\documentclass[fleqn,usenatbib]{mnras}
\usepackage{newtxtext,newtxmath}
\usepackage[T1]{fontenc}
\usepackage{ae,aecompl}
\usepackage{multirow}
\usepackage{subcaption}
\usepackage{enumerate}
\usepackage{graphicx}	% Including figure files
\usepackage{amsmath}	% Advanced maths commands
\title[]{UV-photoprocessing of acetic acid (CH$_3$COOH)-bearing interstellar ice analogs}

\author[C. del Burgo Olivares et al.]{
C. del Burgo Olivares,$^{1}$\thanks{E-mail: cdelburgo@cab.inta-csic.es}
H. Carrascosa, $^{1}$\thanks{E-mail: hcarrascosa@cab.inta-csic.es}
B. Escribano, $^{1}$\thanks{E-mail: bescribano@cab.inta-csic.es}
G. M. Mu\~noz Caro,$^{1}$\thanks{E-mail: munozcg@cab.inta-csic.es}
\newauthor
\;R. Mart\'in-Dom\'enech$^{1}$\thanks{E-mail: rmartin@cab.inta-csic.es}
\\
$^{1}$Centro de Astrobiolog\'{\i}a (CSIC-INTA), Ctra. de Ajalvir, km 4, Torrej\'on de Ardoz, 28850 Madrid, Spain\\
}

\date{Accepted 2023 December 01. Received 2023 November 30; in original form 2023 November 02}

\pubyear{2023}

\begin{document}
\label{firstpage}
\pagerange{\pageref{firstpage}--\pageref{lastpage}}
\maketitle

\begin{abstract}

Acetic acid (CH$_3$COOH) was detected in the gas toward interstellar clouds, hot cores, protostars and comets. Its formation in ice mantles was proposed and acetic acid awaits detection in the infrared spectra of the ice as most other COMs except methanol. The thermal annealing and UV-irradiation of acetic acid in the ice was simulated experimentally in this work under astrophysically relevant conditions. The experiments were performed under ultra-high vacuum conditions. An ice layer was formed by vapor deposition onto a cold substrate, and was warmed up or exposed to  UV photons. The ice was monitored by infrared spectroscopy while the molecules desorbing to the gas phase were measured using a quadrupole mass spectrometer. The transformation of the CH$_3$COOH monomers to cyclic dimers occurs at 120 K and the crystal form composed of chain polymers was observed above 160 K during warm-up of the ice. Ice sublimation proceeds at 189 K in our experiments. Upon UV-irradiation simpler species and radicals are formed, which lead to a residue made of complex molecules after warm-up to room temperature. The possible formation of oxalic acid needs to be confirmed. The photodestruction of acetic acid molecules is reduced when mixed with water in the ice. This work may serve to search for the acetic acid photoproducts in lines of sight where this species is detected. A comparison of the reported laboratory infrared spectra with current JWST observations allows to detect or set upper limits on the CH$_3$COOH abundances in interstellar and circumstellar ice mantles.\\

\end{abstract}


%%%%%%%%%%%%%%%%%%%%%%%%%%%%%%%%%%%%%%%%%%%%%%%%%% 
%%%%%%%%%%%%%%%%% BODY OF PAPER %%%%%%%%%%%%%%%%%%
%%%%%%%%%%%%%%%%%%%%%%%%%%%%%%%%%%%SECTION INTRODUCCIÓN%%%%%%%%%%%%%%%%%%%%%%%%%%%%%%%%%%%%%%
\section{Introduction}
\label{sect.introduction}

In the interior of dense interstellar clouds, the particle density of 10$^{3}$-10$^{6}$ particles cm$^{-3}$ prevents the penetration of ultraviolet (UV) radiation emitted by massive stars, reaching temperatures below 20 K and leading to the accumulation of volatile substances on the surface of dust grains forming an ice layer. This interstellar ice is mainly composed of H$_2$O, CO, CO$_2$, CH$_3$OH and NH$_3$. Their exposure to secondary UV-photons produced by cosmic-ray induced hydrogen excitation in the cloud, and the processes taking place when grains experience heating in hot cores lead to an increase in the chemical complexity of the ice molecular composition. Up to now, more than 90 complex organic molecules (COMs) have been detected in the interstellar medium (ISM) by astronomical observations \citep[]{McGuire2022}. A significant fraction of these molecules are expected to have been produced in the icy mantles. Some of these molecules, such as carboxylic acids (R-COOH), have received considerable interest due to their structural similarity to amino acids and other biologically important species. In the astrobiological context, acetic acid (CH$_3$COOH) acquires significant relevance as a key precursor of glycine (NH$_2$CH$_2$COOH). Its carboxylic structure allows direct formation by replacing a hydrogen atom with an amino group (-NH$_2$). Theoretical \citep[]{Garrod2013, Sato2018} and experimental \citep[]{Blagojevic2003, jackson2005, Lafosse2006} studies have reported the formation of glycine from acetic acid, both in the gas phase and in interstellar ice analogs.\\

The first detection of acetic acid in the interstellar medium (ISM) was towards Sagittarius B2 Large Molecule Heimat (Sgr B2 LMH) \citep[]{Mehringer1997}. It was subsequently detected in the hot-core source W51e2 \citep[]{Remijan2002}, in the hot high-mass molecular cloud G34.3 + 0.2 \citep[]{Remijan2003}, in the low-mass protostars IRAS 16293-2422 \citep[]{Cazaux2003} and G19.61-0.23 \citep[]{Shiao2010}, in the nebula CRL 618 \citep[]{Remijan2005} and Orion-KL \citep[]{Favre2017}, and towards massive star-forming regions MM1 and MM2 in the NGC 6334 complex \citep[]{ElAbd2019}. In the solar system, the ESA-Rosetta mission detected acetic acid ejected from the surface of comet 67P/Churyumov-Gerasimenko \citep[]{Altwegg2017, Schuhmann2019} at concentrations of 0.0034\% with respect to H$_2$O. For comet C/1996 O1 (Hale-Bopp) upper limits of 0.06\% were proposed \citep[]{Crovisier2004}. The column densities of CH$_3$COOH range from 8 $\times$ 10$^{14}$ to 3  $\times$ 10$^{17}$ cm$^{-2}$ toward 15 star-forming regions analysed by \cite{ElAbd2019}.\\

The relative abundance of acetic acid in interstellar/circumstellar ice mantles is not known. Among other species related to acetic acid, the observed IR absorption band at 7.24 $\mu$m (1381 cm$^{-1}$) may be attributed to the C-H deformation mode of formic acid (HCOOH), while the 7.41 $\mu$m (1350 cm$^{-1}$) band is compatible with the formate ion (HCOO$^-$), although acetaldehyde (CH$_3$HCO) is also used for the fit of this band \citep[]{Schutte1999}.\\

Formation of acetic acid in the interstellar medium has been studied by many authors. \cite{Huntress1979} proposed a mechanism of acetic acid formation in the gas phase based on radiative association (CH$_3$CO$^{+}$ + H$_2$O → CH$_3$COOH$_2$$^{+}$) followed by dissociative recombination (CH$_3$COOH$_2$$^{+}$ + e$^{-}$ → CH$_3$COOH + H).  \cite{Ehrenfreund2000} proposed this mechanism to explain the formation of acetic acid from the reaction between protonated methanol (CH$_3$OH$_2$$^{+}$) and formic acid (HCOOH) evaporated from grain surfaces at high temperature. However, observations by \cite{Mehringer1997} and \cite{Remijan2002} of acetic acid in Sgr B2 suggested that gas-phase reactions alone are not sufficient to explain the formation of this species in these environments, and therefore processes occurring within interstellar ice mantles become important.  \cite{Bennett_Kaiser2007} and  \cite{Bergantini2018} suggested that acetic acid can be formed through radical-radical recombination of methyl (CH$_3$) with hydroxycarbonyl (HOCO) in apolar ices containing carbon dioxide (CO$_2$) and methane (CH$_4$), exposed to ionising radiation from energetic electrons. \cite{Kleimeier2020} deduced that the formation of acetic acid in polar interstellar ices (D$_2$O: CH$_3$CHO), irradiated by energetic electrons, occurs through the recombination of acetyl (CH$_3$CO) and hydroxyl-d$_1$ (OD) radicals. Quantum chemistry studies \citep[]{Ahmad2020} propose an alternative pathway via the reaction between two formaldehyde (HCHO) molecules. The results reveal that the formation of acetic acid is more favourable in icy grains than in the gas phase, suggesting bond formation through the catalytic activity of the dust grains.\\

In recent years, several studies have been carried out on the thermal adsorption and desorption of CH$_3$COOH-bearing ices on surfaces analogous to interstellar dust grains \citep[]{Bahr&Borodin2006,Bertin2011,Lattelais2011, Burke2015}, providing information on the structural changes that CH$_3$COOH molecules undergo with temperature. Similarly, the processing of acetic acid ices under ionising radiation has been studied. \cite{Bernstein2004} revealed the stability of acetic acid in the ice to UV-photolytic destruction. \cite{Macoas2004} reported the decomposition pathways of CH$_3$COOH to the photoproducts detected in an Ar matrix. Irradiation of acetic acid ice samples by soft X-rays allowed to detect the most abundant ionic fragments (CH$_3$CO$^+$, COOH$^+$, H$^+$, y CH$_3$$^+$) \citep[]{Pilling2006} as well as to quantify the main products formed in the photodissociation of CH$_3$COOH \citep[]{RACHID2017}.
Despite the available studies, several questions are still open about the role of acetic acid as a precursor of more chemically complex species. In this work, we performed laboratory experiments with ice samples containing acetic acid in three different environments. Irradiation of these low-temperature ices with UV photons and subsequent warm-up to room temperature serve to mimic the chemical processes involving acetic acid in the ice under insterstellar and circumstellar conditions.\\
 
This paper is structured as follows. Sect. \ref{sect.experimental_setup} describes the experimental methods and the DFT simulations used in this work. Sect. \ref{Results} reports the results extracted from the experiments. Sect. \ref{sect.enviroments} presents the IR features of acetic acid ice samples in different environments and a comparison with numerical simulations. Sect. \ref{sect.UV_irradiation} focuses on the photoproducts quantification after irradiation of the samples at 10 K. Sect. \ref{sect.warmup} addresses the formation of complex species after warm-up of the irradiated ice samples. Conclusions and astrophysical implications regarding this work are presented in Sects. \ref{sect.conclusions} and \ref{sect.astrophysical_implications}.\\

%%%%%%%%%%%%%%%%%%%%%%%%%%%%%%%%SECCIÓN EXPERIMENTAL%%%%%%%%%%%%%%%%%%%%%%%%%%%%%%%%%%%%%%%%%
\section{ExperimentaL}
\label{sect.experimental_setup}

\begin{table*}%Experimentos Realizados
\small
   %\centering
\caption[]{Parameters of the experiments performed.}
    \label{Table.experiments}
    %\resizebox{9cm}{!}{
    %\begin{flushleft}
    \begin{tabular}{cccccc}
\hline
Exp. & Comment  & Ice composition  & N$_{0}$(CH$_3$COOH)   & Photon dose   & Heating rate \\
           &          &             & (cm$^{-2}$)       & (cm$^{-2}$)   & (K/min)\\
\hline
\noalign{\smallskip}
\textbf{1} & Deposition & CH$_3$COOH  &  $6.1\times10^{17}$& 0    &0.5 \\
\noalign{\smallskip}
\textbf{2} & Deposition & CH$_3$COOH &  $5.7\times10^{17}$ & 0    &1 \\
\noalign{\smallskip}
\textbf{3} & Deposition$^{\text{a}}$ & CH$_3$COOH &  $3.9\times10^{17}$ &0     &1 \\
\noalign{\smallskip}
\textbf{4} & Deposition$^{\text{b}}$ & CH$_3$COOH &  $5.4\times10^{17}$ & 0         &1   \\%Dep. 160 K
\noalign{\smallskip}
\textbf{5} & Dep. \& Irrad. & CH$_3$COOH & $4.9\times10^{17}$  & $1.2\times10^{18}$   &1   \\
\noalign{\smallskip}
\textbf{6} & Dep. \& Irrad. & CH$_3$COOH  &  $5.8\times10^{17}$ & $1.1\times10^{18}$   &0.5 \\
\noalign{\smallskip}
\textbf{7} & Dep. \& Irrad. & CH$_3$COOH &  $4.8\times10^{17}$ &$1.2\times10^{18}$    &0.5 \\
\noalign{\smallskip}
\textbf{8} & Dep. during irrad. & CH$_3$COOH &  -& $1.7\times10^{18}$   &0.3 \\
\noalign{\smallskip}
\textbf{9} & Dep. \& Irrad. & CH$_3$COOH:Xe &  $2.3\times10^{17}$ & $2.3\times10^{18}$   &0.5 \\
\noalign{\smallskip}
\textbf{10} & Deposition & CH$_3$COOH:H$_2$O (1:1) &  $2.7\times10^{17}$ &0     &0.5 \\
\noalign{\smallskip}
\textbf{11} & Dep. \& Irrad. & CH$_3$COOH:H$_2$O (1:0.6) &  $4.9\times10^{17}$ &$4.5\times10^{17}$    &0.5 \\
\noalign{\smallskip}
\textbf{12} & Dep. \& Irrad. & CH$_3$COOH:H$_2$O (1:1.6) &  $2.4\times10^{17}$ &$8.5\times10^{17}$     &0.5 \\
\noalign{\smallskip}
\textbf{13} & Dep. \& Irrad. & CH$_3$COOH:H$_2$O (1:2.5) &  $2.5\times10^{17}$ &$5.5\times10^{18}$    &0.5 \\
\hline
\end{tabular}\\
%\end{flushleft}
%}\\
%\begin{center}
\begin{flushleft}
\textit{Deposition temperature: $^{\text{a}}$ 160 K, $^{\text{b}}$ 120 K, all other experiments: 10 K.}\\
\end{flushleft}
%\small{}\\ 
\end{table*}% Experimentos Realizados

\subsection{ISAC experimental simulations}%Sección experimental de la cámara

Experiments were carried out in the InterStellar Astrochemistry Chamber (ISAC) located at Centro de Astrobiología, described in more detail in \cite{MuñozCaro2010}. ISAC is an ultra-high vacuum chamber with a base pressure of 4 x 10$^{-11}$ mbar, a pressure similar to the one in the interior of dense interstellar clouds. ISAC is equipped with a closed-cycle helium cryostat able to reach temperatures of 8 K on the KBr substrate used for ice deposition.\\

The chemical components used for these experiments were distilled MilliQ H$_2$O, obtained from Millipore; CH$_3$COOH $\geq$ 99~\% from Sigma-Aldrich and Xe acquired from Praxair with 99.999~\% purity.\\

ISAC gas line system has a quadrupole mass spectrometer (QMS, Pfeiffer Vacuum, Prisma QMS 200) which allows to control the molecular composition before entering the chamber and to detect any trace of contamination. Gases and vapours were introduced into the chamber at a pressure of 2 x 10$^{-7}$ mbar through a deposition tube oriented towards the substrate. During deposition, irradiation, and thermal processing of the samples, the gas phase was monitored in the main chamber using another QMS (Pfeiffer Vacuum, Prisma QMS 200) equipped with a Channeltron detector. Solid phase was monitored by Fourier Transform Infrared Spectroscopy (FTIR) in transmittance mode using a Bruker Vertex 70 at a resolution of 1 or 2 cm$^{-1}$. A reference IR spectrum was measured before deposition to allow subtraction of the bare window. IR spectra were also acquired after deposition, after each irradiation interval and during warm-up of the ice. Column density of each species was calculated using the equation \ref{Eq.Band.strength}:\\

\begin{equation}
 N = \frac{1}{A} \int_{\text{band}}{\tau _{v}\; dv.} 
 \label{Eq.Band.strength}
\end{equation}\\ %Ecuación de la fuerza de banda

where $N$ is the column density in cm$^{-2}$, $\tau_{\nu}$ the optical depth of the band, $d\nu$ the wavenumber differential in cm$^{-1}$, and $A$ the band strength in cm molecule$^{-1}$. For pure acetic acid, the band strength reported by \cite{Bennett_Kaiser2007}, $5.2 \times 10^{-17}$ cm molecule$^{-1}$ for the C=O stretching mode centered around 1750 cm$^{-1}$, was adopted. For binary CH$_3$COOH:H$_2$O ice mixtures, the C=O stretching mode of CH$_3$COOH overlaps with the band at 1613 cm$^{-1}$ assigned to H$_2$O. Therefore, a band strength of  
$2.6 \times 10^{-18}$ cm molecule$^{-1}$ was calculated from pure CH$_3$COOH ices for the rocking mode of CH$_3$ at 1020 cm$^{-1}$. For H$_2$O, the value of $2.0 \times 10^{-16}$ cm molecule$^{-1}$ for the 3280 cm$^{-1}$ band was used \citep{HAGEN1981}. For CO, we used $8.7 \times 10^{-18}$ cm molecule$^{-1}$ for the C$\equiv$O stretching mode at 2136 cm$^{-1}$ \citep{GonzalezDiaz2022}, and for CO$_2$, $1.3 \times 10^{-16}$ cm molecule$^{-1}$ for the C=O stretching mode centered at 2343 cm$^{-1}$ \citep{Bouilloud2015}.\\

Ice samples were irradiated at 10 K using an F-type microwave-discharged hydrogen flow lamp (MDHL), from Opthos Instruments. The Evenson cavity of the lamp is refrigerated with dry air. This lamp provides an average UV flux of $1.3 \times 10^{14}$ cm$^{-2}$ s$^{-1}$ at the sample position for a hydrogen flux of 0.4 mbar, as measured with a calibrated Ni-mesh placed at the end of the MDHL. MDHL emission, reported in \cite{Cruz-Diaz2014}, is located in the 114-180 nm (10.87 - 6.89 eV) range. The main emission bands are Ly-$\alpha$ at 121.6 nm (10.20 eV) and the molecular hydrogen bands at 157.8 nm (7.85 eV) and 160.8 nm (7.71 eV). The high VUV absorbance of the KBr substrate inhibits VUV photon transmission. For this reason, a MgF$_2$ substrate was used for vacuum ultraviolet spectroscopy measurements. VUV spectrum was recorded using a McPherson 0.2 m focal lenght VUV monochromator (model 234/302). The spectrum recorded by the VUV detector, located at the rear of the chamber, is attenuated by three MgF$_2$ windows (absorbing most of the Ly-$\alpha$ emission). Therefore, VUV spectrum was measured between 130 - 180 nm (9.54 - 6.89 eV) in the reported experiments. Nonetheless, the UV emission spectrum corresponding to a single MgF$_2$ window absorption \citep{Cruz-Diaz2014} is the one experienced by the ice sample during irradiation. The VUV absorption spectrum of CH$_3$COOH ice samples was obtained after substraction of the absorbance spectrum of the neat MgF$_2$ window. VUV absorption cross-section was estimated following \cite{Cruz-Diaz2014}, using Eq. \ref{Eq.VUV_absorption.}, based on Lambert-Beer law,

\begin{equation} %VUV-absorption Cross-section
I _{\text{t}}(\text{$\lambda$}) = I _{\text{0}}(\text{$\lambda$}) \cdot e^{- \sigma(\text{$\lambda$}) \cdot N}
\label{Eq.VUV_absorption.}
\end{equation}\\%VUV-absorption Cross-section

where $I_{\text{t}}(\text{$\lambda$})$ is the transmitted intensity for a given wavelength $\lambda$, $I_{\text{0}}(\text{$\lambda$})$ is the incident intensity, $\sigma(\text{$\lambda$})$ is the VUV-absorption cross-section of the ice in cm$^{2}$ and $N$ is the average of the column density before and after irradiation in cm$^{-2}$.\\

Quantitative information on the photodestruction efficiency of CH$_3$COOH can be obtained from the UV destruction cross-section of CH$_3$COOH ice, which can be calculated using Eq. \ref{Eq.photodestruction.},\\

\begin{equation} %Ecuación de la sección eficaz de fotodestrucción
N(\text{t}) = N(\text{0}) \cdot e^{- \phi \cdot t \cdot \sigma _{\text{des}}}
\label{Eq.photodestruction.}
\end{equation}\\%Ecuación de la sección eficaz de fotodestrucción

where $N(t)$ and $N(0)$ are the column densities (cm$^{-2}$) before and after irradiation, respectively, $\phi$ is the UV flux (cm$^{-2}$ s$^{-1}$), $t$ is the irradiation time (s), and $\sigma_{\text{des}}$ is the destruction cross-section (cm$^2$) \citep[]{Carrascosa2020}, related to the first irradiation period before recombination reactions of CH$_3$COOH become important.\\

Furthermore, formation cross-sections ($\sigma_{\text{form}}$) for the main photoproducts were obtained using equation \ref{Eq.photoformation},

\begin{equation} %Ecuación de la sección eficaz de formación
\centering
\frac{\mathrm{d}N}{\mathrm{d}t} = \phi \cdot t \cdot N(\text{CH$_3$COOH}) \cdot \sigma _{\text{form}}
\label{Eq.photoformation}
\end{equation}\\ %Ecuación de la sección eficaz de formación

where $\frac{dN}{dt}$ is the variation of the column density of the photoproducts during the irradiation period, $t$ is the irradiation time, and $N(\text{CH$_3$COOH})$ is the average column density of acetic acid during the irradiation interval. For long irradiation periods, some photoproducts may enhance the formation of the species of interest. Therefore, the first irradiation period was taken for calculation of the formation cross-sections from acetic acid ice samples.\\

Photodesorption of species during UV irradiation of the ice samples was quantified using Eq. \ref{Eq.fotodesorption},

\begin{equation}
\centering
N(\text{mol}) = \frac{A(\text{m/z})}{k_{\text{CO}}} \cdot \frac{\sigma^+(\text{CO})}{\sigma^+(\text{mol})} \cdot \frac {I_ {F}\left (\mathrm {CO} ^ {+}\right)} {I_ {F} (z)} \cdot \frac {F_ {F} (28)} {F_ {F} (m)} \cdot \frac {S (28)} {S (m/z)}
\label{Eq.fotodesorption}
\end{equation}\\ %Ecuación de moléculas que fotodesorben

where $N$(mol) is the total number of photon-induced desorbed molecules, $A$ (m/z) is the integrated area of the QMS signal during photon-induced desorption, $k_{\text{CO}}$ is the proportionality constant of the QMS calibration in a CO ice irradiation experiment \citep[][8.33 $\times$ 10$^{-26}$ A min cm$^{2}$ molecule$^{-1}$ at the time the experiments were performed]{Martin-Domenech2015}, $\sigma^+$ (mol) is the ionization cross section of the ionized species at a voltage of 70 eV in the QMS adopted from NIST, $I_F(z)$ is the ionization factor, which is considered to be 1 for all molecules, $F_F(m)$ is the fragmentation factor, obtained from the QMS spectrum of each species, and $S$ (m/z) is the QMS sensitivity, which depends on the QMS current as a function of the mass of each molecule (see \cite{Martin-Domenech2015} for calculation details). After irradiation, ice samples were warmed-up to room temperature using a Lakeshore temperature controller 331, allowing temperature programmed desorption (TPD) experiments, with an accuracy better than 0.1 K.\\

\subsection{IR spectra simulations}%Simulaciones de Bruno
With the purpose of assigning vibrational bands and supporting the experimental results, we performed density functional theory (DFT) \citep{DFT-HK,DFT-KS} simulations using the CASTEP software \citep{CASTEP}.
We used the generalized gradient approximation (GGA) and the Perdew-Burke-Ernzerhof functionals \citep{PBE1996}.
During geometry optimization we selected a tolerance of 10$^{-5}$ eV/atom for the energy. The force tolerance parameter was set to 0.03 eV/\AA, and the maximum stress
was set to 0.05 GPa.
Simulated spectra were computed using density-functional perturbation theory \citep{CASTEP_DFPT} with a convergence tolerance of 10$^{-5}$\AA.
Initial crystallographic structures for acetic acid were taken from \cite{Boese1999}.
Amorphous solids were generated by melting the crystal using molecular dynamics in the NPT ensemble at 200 K for 1 ps. We used the Nose-Hoover thermostat \citep{Nose1985} and the Andersen barostat \citep{Andersen1980}.\\

%%%%%%%%%%%%%%%%%%%%%%%%%%%%%%%%RESULTADOS%%%%%%%%%%%%%%%%%%%%%%%%%%%%%%%%%%%%%%%%%%%%%%%%%%%%%%%%%%%%%%%%%%%%%%%%%%%%%%%%%%%%%%%%%%%%%%%%%%%%%%%%%%%%%%%%%%%%%
%%%%%%%%%%%%%%%%%%%%%SECCION CH3COOH NO IRRADIADO%%%%%%%%%%%%%%%%%%%%%%%%%%%%%%%%%%%%%%%
\section{Results and discussion}
\label{Results}

Parameters for each of the experiments presented in this work are compiled in Table \ref{Table.experiments}. The first three columns indicate the label of the experiment, if the experiment involves irradiation and the initial composition of the ice, respectively. The fourth column indicates the initial column density of CH$_3$COOH in the ice. The fifth column gives the irradiation dose used and the sixth column the heating rate of the ice.\\

\subsection{Effect of ice environment on crystallization of CH$_3$COOH ice samples.}
\label{sect.enviroments}

\begin{figure*}
  \centering 
  \includegraphics[width=1\textwidth]{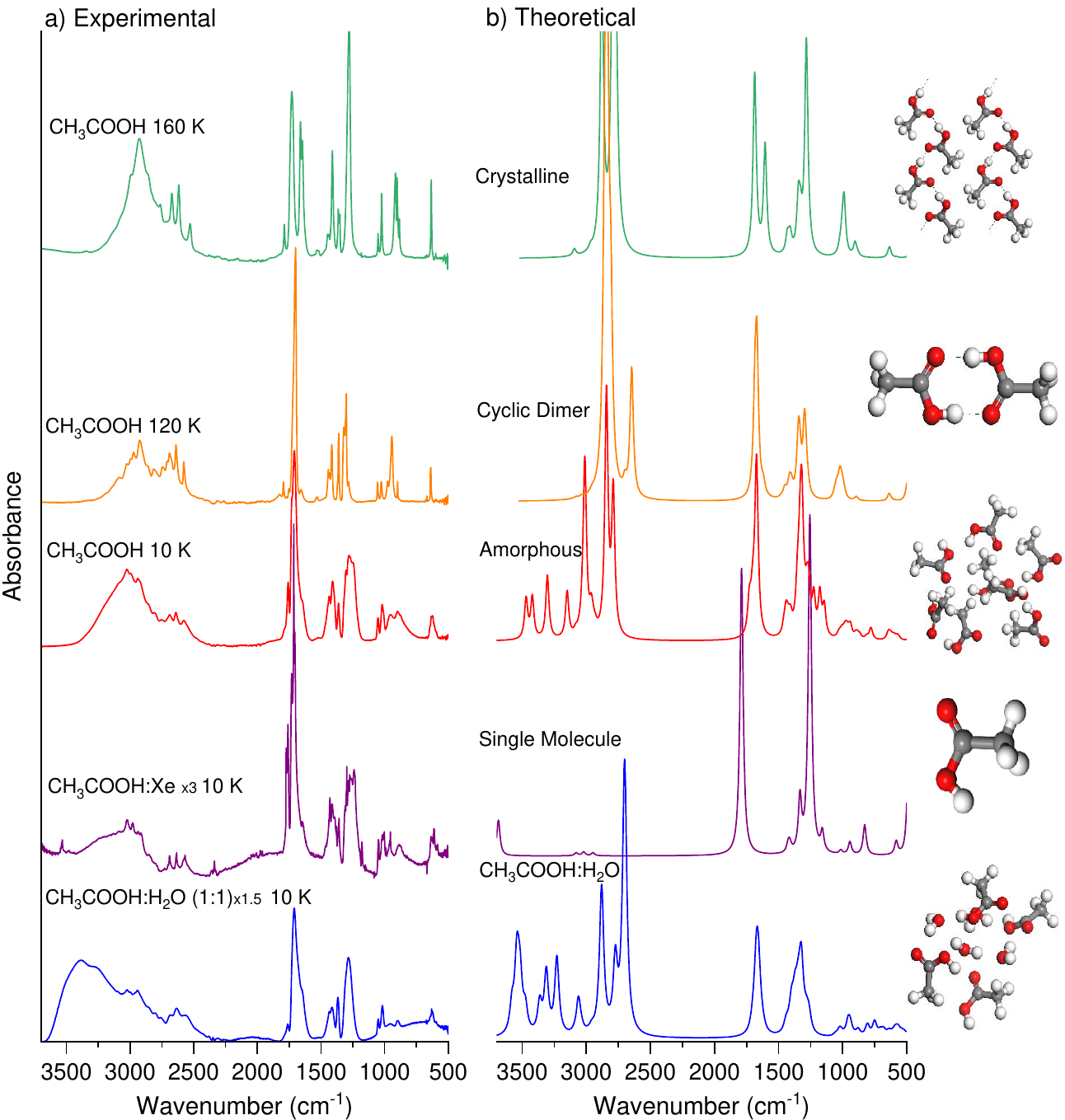}
  \caption{\textbf{a)} Experimental and \textbf{b)} DFT-calculated IR spectra of acetic acid ice samples in several environments (in order from top to bottom: experiments \textbf{4}, \textbf{3}, \textbf{1}, \textbf{9} and \textbf{10}). The absorbance of CH$_3$COOH:Xe and CH$_3$COOH:H$_2$O experimental spectra has been expanded for a better comparison between spectra.}
  \label{Fig.HAc_IR_diferentesEntornos}
\end{figure*}

Table \ref{Table.IR_sinirradiar} shows the position of the IR features of acetic acid after deposition of pure CH$_3$COOH (Experiment \textbf{1}), CH$_3$COOH in a Xe-matrix (Exp. \textbf{9})  and in a 1:1 mixture with water (Exp. \textbf{10}). Experimental IR spectra have been accompanied by theoretical DFT simulations to facilitate the identification of the individual vibrational bands (Fig. \ref{Fig.HAc_IR_diferentesEntornos}). The main effect of the Xe-matrix is the reduction of the intermolecular forces, as in a single molecule model, leading to sharper IR bands and the appearance of a new narrow band at 3535 cm$^{-1}$, associated with the O-H stretching of isolated CH$_3$COOH molecules. With thermal processing, pure acetic acid ice samples undergo structural changes that can be observed through changes in the C=O and C-O vibration modes. When acetic acid is deposited at 10 K, C=O stretching vibrations are observed around 1710 cm$^{-1}$, related to the organisation of cyclic dimers formed in the gas phase, and at 1758 cm$^{-1}$, related to the amorphous arrangement of the monomers trapped in the bulk, while C-O stretching vibrations are found around 1280 cm$^{-1}$ \citep[]{Bertin2011}. Thermal annealing to 120 K causes the disappearance of the amorphous monomers and their transformation to cyclic dimers, as suggested by the increase in the intensity of the feature at 1710 cm$^{-1}$. At larger temperatures, around 160 K, new C=O stretching vibrations are observed at 1660 cm$^{-1}$ and 1648 cm$^{-1}$, as well as the narrowing of the C-O stretching band near 1280 cm$^{-1}$. These changes in the features are indicative of the phase transition of acetic acid from a cyclic dimer organisation to a crystalline structure in the form of chain polymers \citep[]{Bertin2011}. The CH$_3$COOH:H$_2$O mixture (1:1) shows a generalised broadening of the bands by interactions with O-H groups. The position of the IR bands in the CH$_3$COOH:H$_2$O mixture are close to the ones measured for the pure ice, see Table \ref{Table.IR_sinirradiar}. \\\

\begin{figure}
  \centering 
  \includegraphics[width=\linewidth]{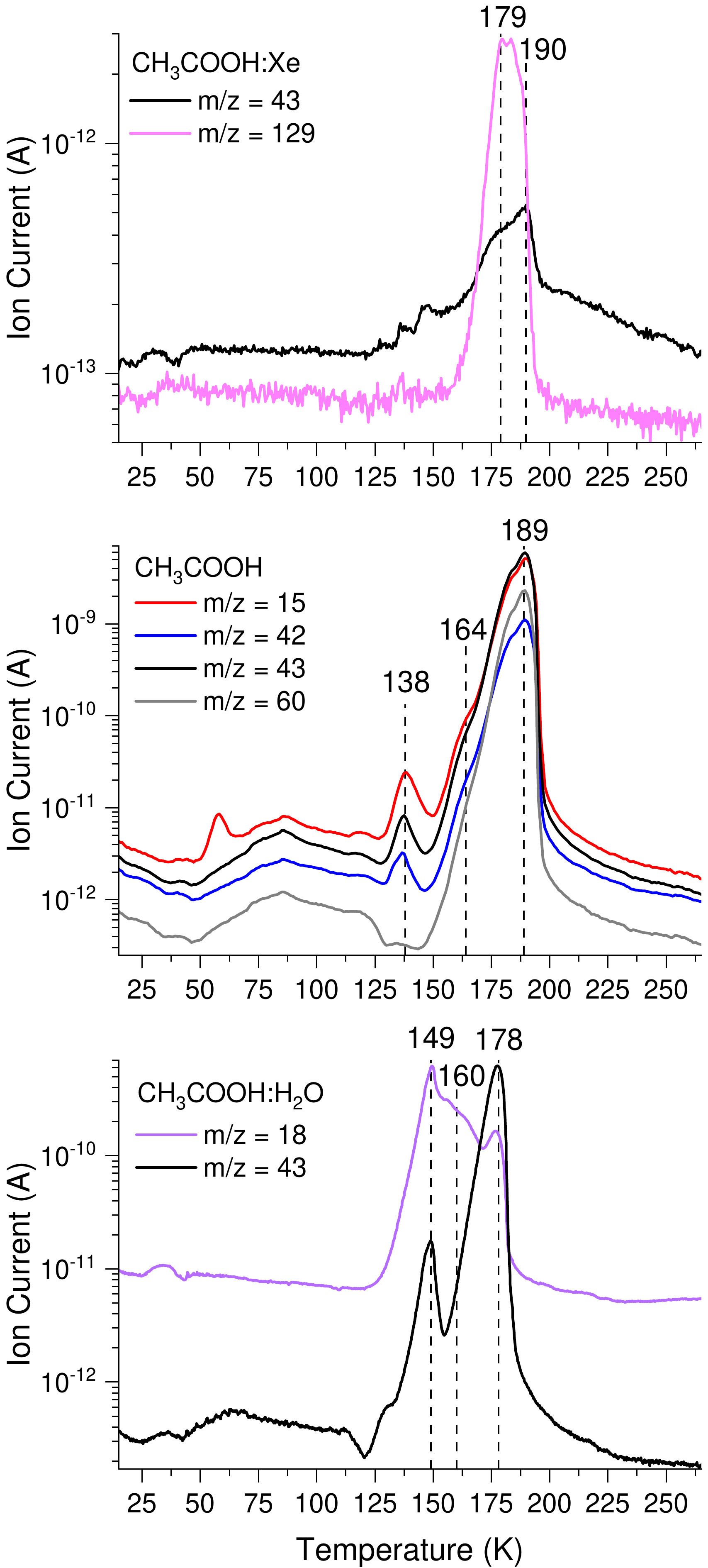}
  \caption{TPD curves of non-irradiated CH$_3$COOH:Xe (Exp. \textbf{9}), CH$_3$COOH (Exp. \textbf{2}) and  CH$_3$COOH:H$_2$O (Exp. \textbf{10}) ice samples recorded by QMS.}
  \label{Fig.TPD_Diferentes_Entornos} 
\end{figure}

\begin{figure}
  \centering
  \includegraphics[width=0.48\textwidth]{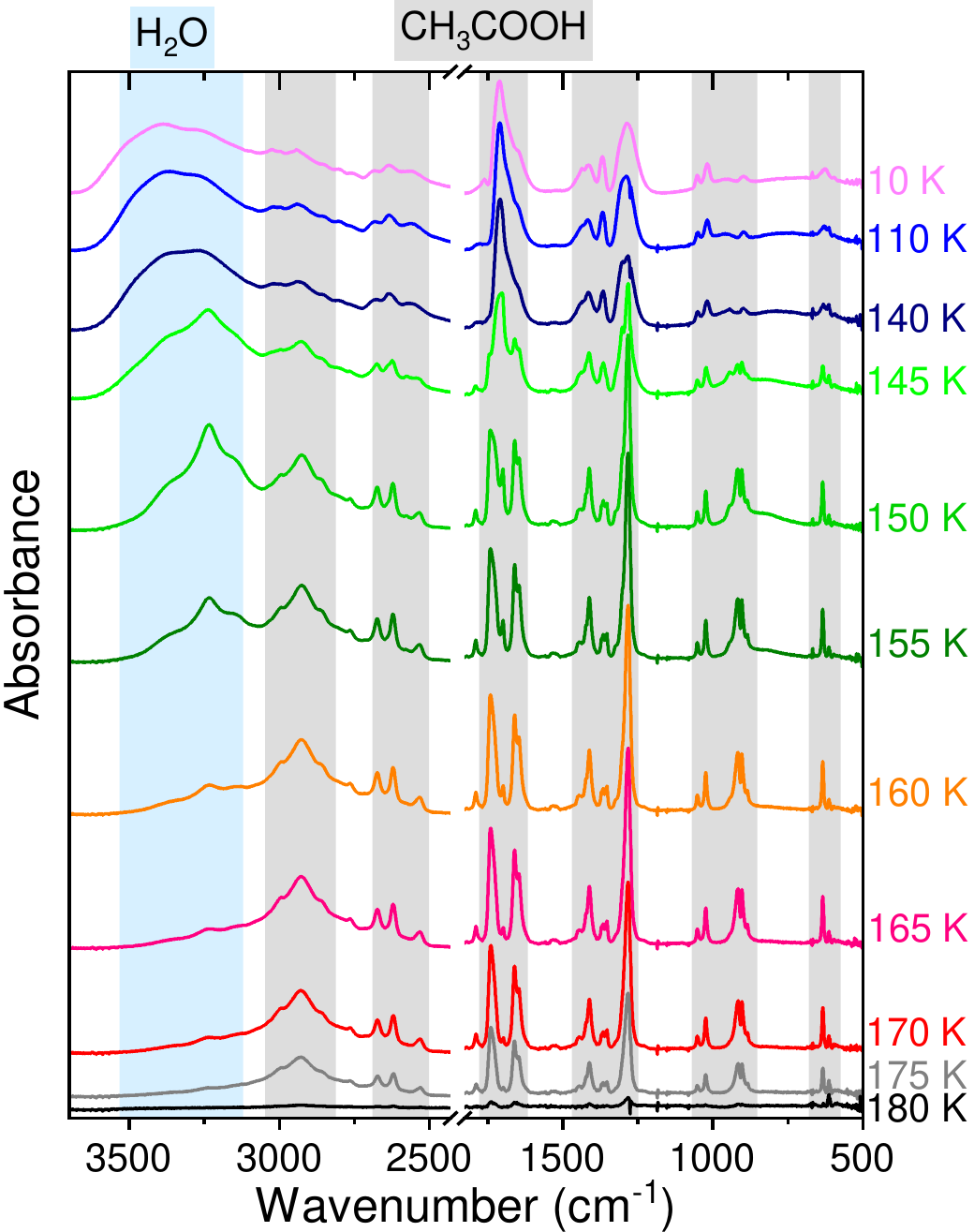}
  \caption{IR spectra during warm-up of non-irradiated CH$_3$COOH:H$_2$O ice mixture (Exp. \textbf{10}). To highlight the behaviour of each compound in ice, blue column corresponds to the main IR features of water and the grey columns to those of acetic acid.}
  \label{Fig.IR_desorción_H2O_HAc}
\end{figure}

Figure \ref{Fig.TPD_Diferentes_Entornos} shows thermal desorption of CH$_3$COOH in the different ice composition studied in this work (Experiments \textbf{9}, \textbf{2} and \textbf{10}) . Pure CH$_3$COOH has a sublimation maximum at 189 K, with previous minor desorptions near 138 and 164 K corresponding to molecules ejected to the gas during phase transitions mentioned above. In the case of CH$_3$COOH:H$_2$O, comparing TPD curves with the evolution of the IR spectra during the warm-up of the ice sample (Fig. \ref{Fig.IR_desorción_H2O_HAc}), it can be observed that the maximum at 149 K corresponds to simultaneous crystallisation of CH$_3$COOH and H$_2$O \citep{Bahr&Borodin2006}. At 160--165 K thermal desorption of H$_2$O takes place in this ice mixture, while at 170-180 K desorption of acetic acid occurs. The absence of alterations in the IR bands of CH$_3$COOH and the non-desorption of acetic acid molecules during the desorption of H$_2$O are indicative of the absence of a molecular complex when mixed in our experiments.\\

 % IR HAc crist., HAc dím., HAc am., HAc:Xe, HAc:H2O, 

\begin{table} %Bandas IR puro HAc:Xe vs. HAc puro vs. HAc:H2O 
%\tiny
   %\centering
\caption[]{IR features (in cm$^{-1}$) of CH$_3$COOH in the different ice mixtures at 10 K. Vibrational modes have been assigned by comparing the experimental values with our theoretical calculations (Fig. \ref{Fig.HAc_IR_diferentesEntornos}) and values reported by \cite{Krause_Katon1977}, \cite{Bahr&Borodin2006},  and  \cite{Lopes2010}.}
    \label{Table.IR_sinirradiar}
%    \begin{center}
\resizebox{8.5cm}{!}{
\begin{tabular}{lccc}
\hline
Vibration mode  & CH$_3$COOH:Xe & CH$_3$COOH & CH$_3$COOH:H$_2$O\\
\hline
\noalign{\smallskip}
$\nu$OH              &3535               &-             &-  \\
\noalign{\smallskip}
$\nu$C=O + $\delta$COH              &3085             &3085             &3099  \\
\noalign{\smallskip}
$\nu$OH                               &3024             &3020             &3024  \\
\noalign{\smallskip}
$\nu_{as}$CH$_3$                      &2978             &2993             &2994  \\
\noalign{\smallskip}
$\nu_{s}$CH$_3$                       &2936             &2937             &2942  \\
\noalign{\smallskip}
 2$\delta$COH                         &2851             &2864             &2858  \\
\noalign{\smallskip}
2$\delta_{as}$CH$_3$                  &2804             &2809             &2803  \\
\noalign{\smallskip}
2$\delta_{s}$CH$_3$                   &2729             &2750             &2759  \\
\noalign{\smallskip}
$\nu$C-O + $\delta$COH              &2689             &2687             &2682  \\
\noalign{\smallskip}
$\nu$C-O + $\delta_{s}$CH$_3$       &2635             &2638             &2635  \\
\noalign{\smallskip}
2$\nu$C-O                             &2569             &2576             &2565  \\
\noalign{\smallskip}
$\nu$C=O (M$_S$)                        &1773             &1789             &-  \\
\noalign{\smallskip}
$\nu$C=O (M$_B$)                       &1760             &1758             &1761  \\
\noalign{\smallskip}
$\nu$C=O (D)                          &1731             &1727             &-  \\
\noalign{\smallskip}
$\nu$C=O (D)                          &1714             &1710             &1710  \\
\noalign{\smallskip}
$\nu$C=O                            &1643             &1648             &1650  \\
 \noalign{\smallskip}
$\delta$OCO + $\nu$C-O              &-                &1523             &-  \\
\noalign{\smallskip} 
$\delta$COH                           &1430             &1434             &1435  \\
\noalign{\smallskip}
$\delta_{as}$CH$_3$                   &1412             &1409             &1413  \\
\noalign{\smallskip}
$\delta_{s}$CH$_3$                    &1359             &1362             &1368  \\
\noalign{\smallskip}
$\nu$C-O (D)                          &1310             &1305             &1286  \\
\noalign{\smallskip}
$\nu$C-O                              &1284             &1281             &-  \\
\noalign{\smallskip}
?                                       &1240             &1251             &-  \\
\noalign{\smallskip}
$\rho_{as}$CH$_3$                     &1048             &1053             &1051  \\
\noalign{\smallskip}
$\rho_{s}$CH$_3$                      &1017             &1020             &1018  \\
\noalign{\smallskip}
?                                       &955              &955              &955  \\
\noalign{\smallskip}
$\gamma$OH?                          &855              &898              &898  \\
\noalign{\smallskip}
$\delta$COO                          &613              &624              &632  \\
\noalign{\smallskip}
\hline
\end{tabular}\\ %Bandas IR puro HAc:Xe vs. HAc puro vs. HAc:H2O
%\end{center}
}
%\begin{center}
\textit{$\nu$ = stretching; $\delta$ = bending; $\rho$ = rocking; $\gamma$ = out of plane deformation. M$_S$ = acetic acid monomers located on the surface; M$_B$ = monomers trapped into the bulk; D = dimmers.
}
\end{table}%Bandas IR puro HAc:Xe vs. HAc puro vs. HAc:H2O 

 %TPD HAc, HAc:Xe, HAc:H2O

 % IR desorción HAc:H2O

%%%%%%%%%%%%%%%%%%%%%%%%%%%%%%%%%%%SECCION CH3COOH IRRADIADO%%%%%%%%%%%%%%%%%%%%%%%%%%%%%%%%%%%%%
\subsection{UV irradiation of pure CH$_3$COOH, CH$_3$COOH:Xe and CH$_3$COOH:H$_2$O ice samples.}
\label{sect.UV_irradiation}

\subsubsection{VUV-absorption cross section of acetic acid}
\label{VUV}

\begin{figure} %VUV-absorbtion cross sections
    \centering
    \includegraphics[width=0.48\textwidth]{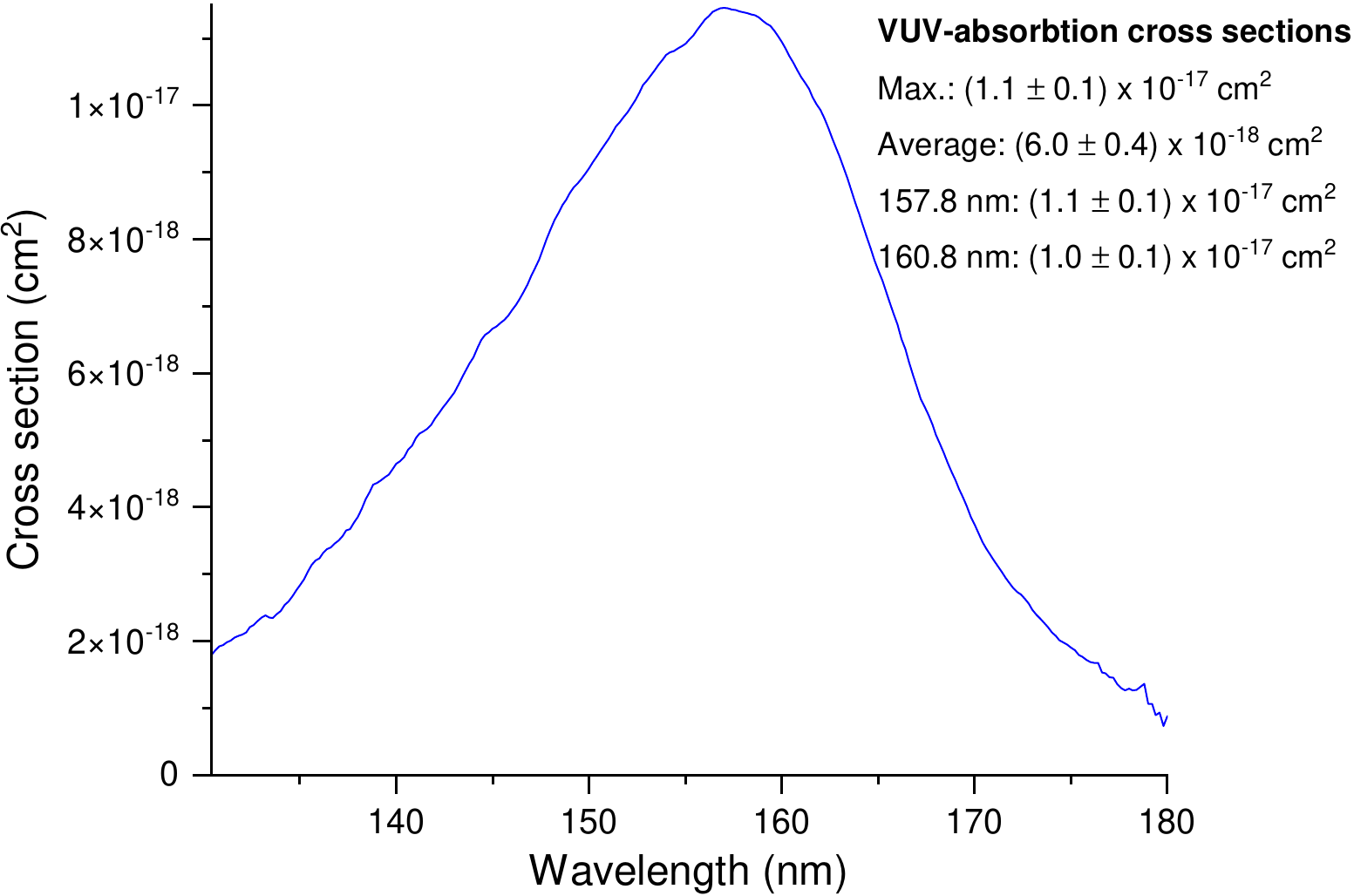}
    \caption{VUV-absorption cross-section of pure CH$_3$COOH ice sample. Max. corresponds to the maximum value of the cross-section in the 130-180 nm range. Average corresponds to the average integrated cross section in the same range. Last two values correspond to the VUV-absorption cross-section at wavelengths 157.8 mm and 160.8 nm, which coincide with the molecular hydrogen bands in the emission spectrum of the MDHL).} 
    \label{Fig.VUVabsorption}
\end{figure} %VUV-absorbtion cross sections

The VUV-absorption cross-section of the ice sample, obtained from Eq. \ref{Eq.VUV_absorption.} allows the calculation of the photon absorption of the ice sample in the 130-180 nm range. VUV-absorption cross-section spectrum of CH$_3$COOH ice at 10 K as a function of wavelength is displayed in Fig. \ref{Fig.VUVabsorption}. It shows a maximum at 157.4 nm (7.88 eV) with a value of (1.1 $\pm$ 0.1) $\times$ 10$^{-17}$ cm$^{2}$. The average VUV-absorption cross-section has a value of (6.0~$\pm$~0.4)~$\times$~10$^{-18}$~cm$^{2}$ in the 130-180 nm (9.54 - 6.89 eV) spectral region. VUV-absorption cross-section of acetic acid ice at the position of the main bands of the Lyman system of H$_2$ near 157.8 nm and 160.8 nm, are (1.1 $\pm$ 0.1) $\times$ 10$^{-17}$ cm$^{2}$ and (1.0 $\pm$ 0.1) $\times$ 10$^{-17}$ cm$^{2}$, respectively. VUV spectroscopy experiments of gas-phase acetic acid carried out by \cite{Leach2006} were performed in the spectral range 61 - 206 nm (20 - 6 eV). The VUV-absorption cross-section of CH$_3$COOH gas is 1.3 $\times$ 10$^{-17}$ cm$^{2}$ at the maximum near 148.5 nm (8.35 eV). At 157.8 and 160.8 nm, \cite{Leach2006} reported values of  1.0 $\times$ 10$^{-17}$ cm$^{2}$ and 9.9 $\times$ 10$^{-18}$ cm$^{2}$, respectively, which are similar to our measurements in the solid phase. \\

\subsubsection{Product identification}
\label{photoproduct}

\begin{figure} %IR de HAc, HAc:Xe y HAc:H2O irradiados
  \centering 
\begin{subfigure}{.38\textwidth}
  \centering 
  \includegraphics[width=\linewidth]{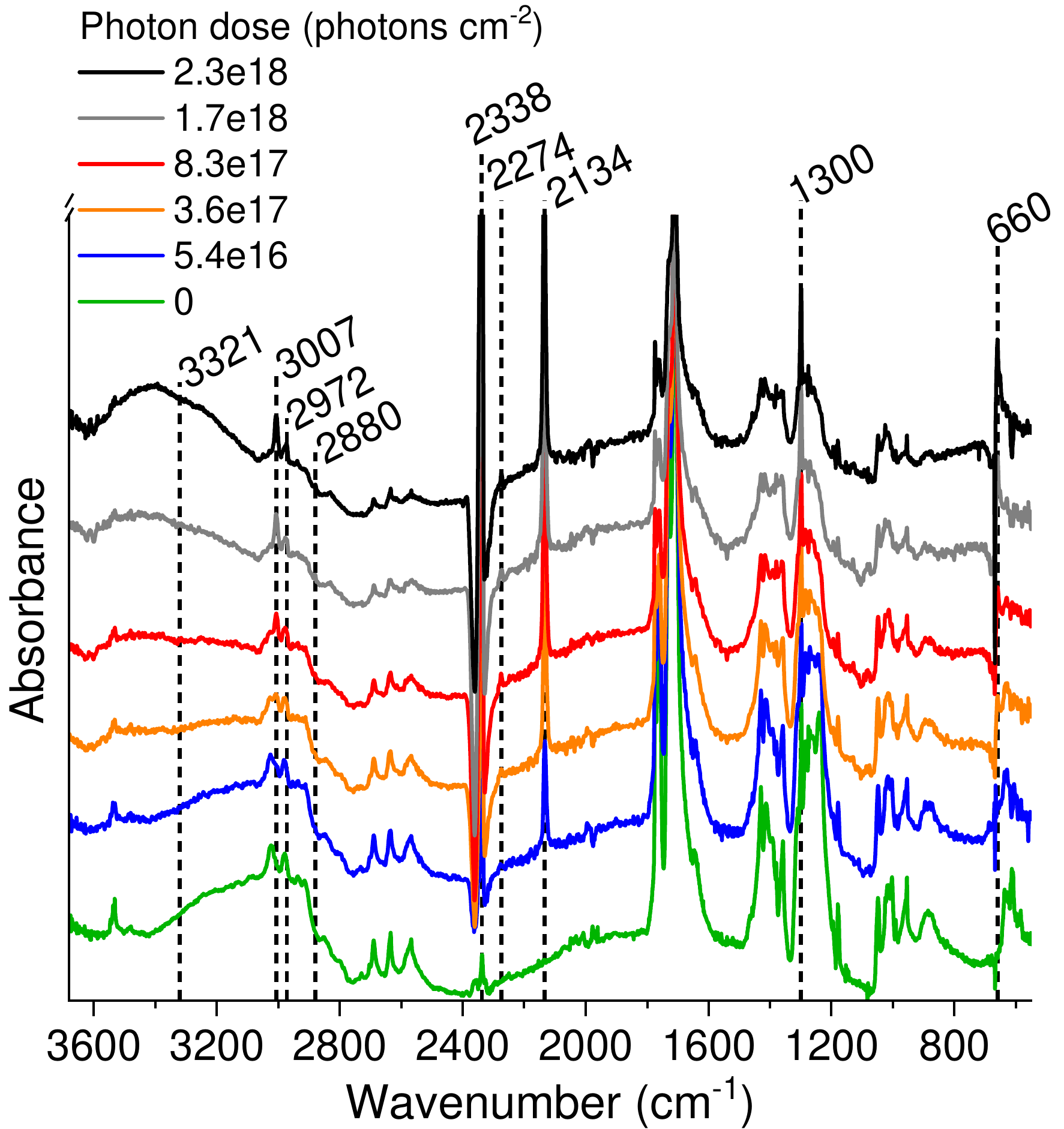}
  \caption{CH$_3$COOH:Xe (Exp. \textbf{9})}
  \label{Fig.HAc-Xe-irradiation.}
\end{subfigure} % irradiación de HAc:Xe
\begin{subfigure}{.38\textwidth}
  \centering 
  \includegraphics[width=\linewidth]{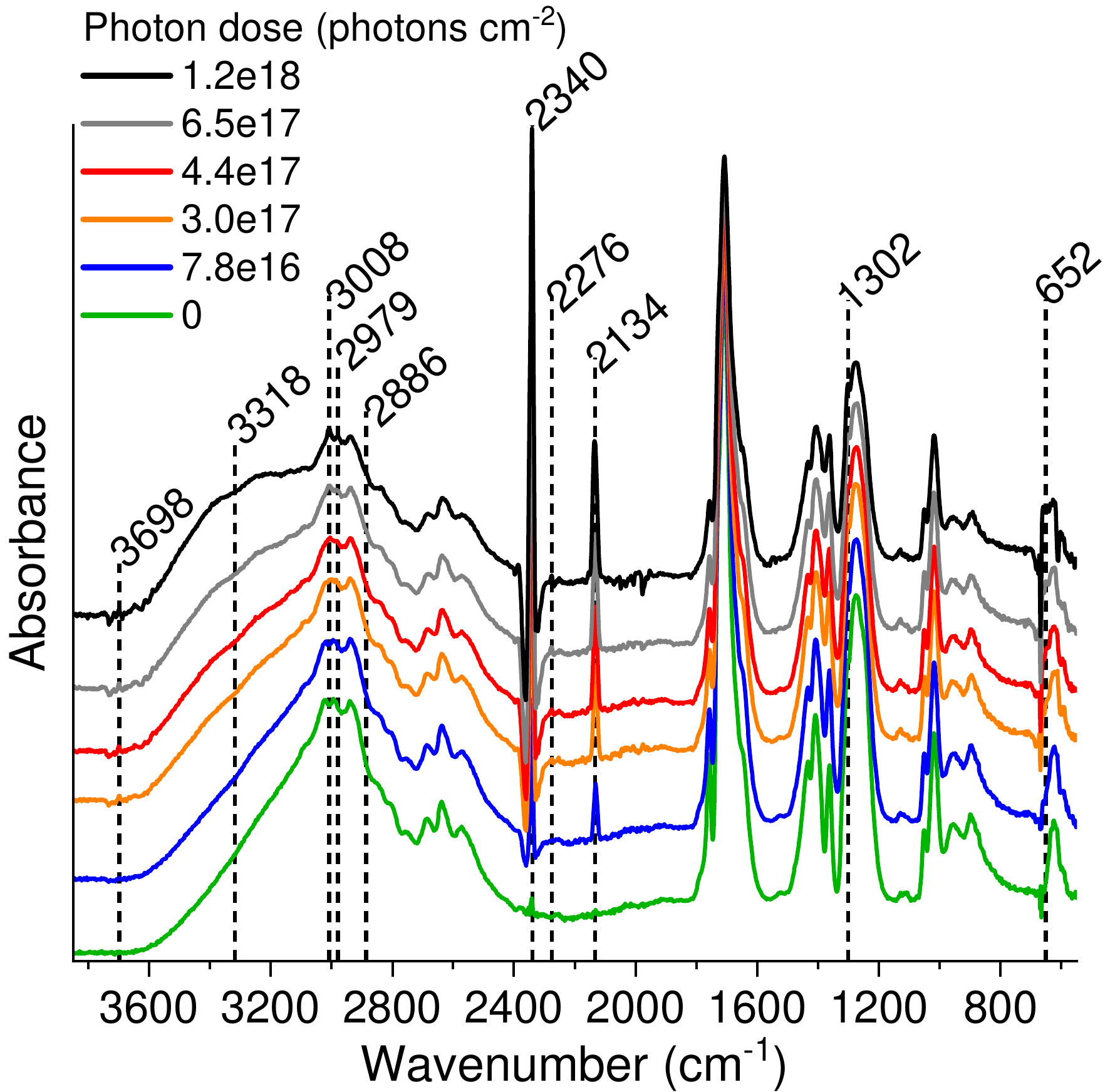}
  \caption{CH$_3$COOH (Exp. \textbf{5})}
  \label{Fig.HAc_irradiation.}
\end{subfigure} %irradiación de HAc
\begin{subfigure}{.38\textwidth}
  \centering 
  \includegraphics[width=\linewidth]{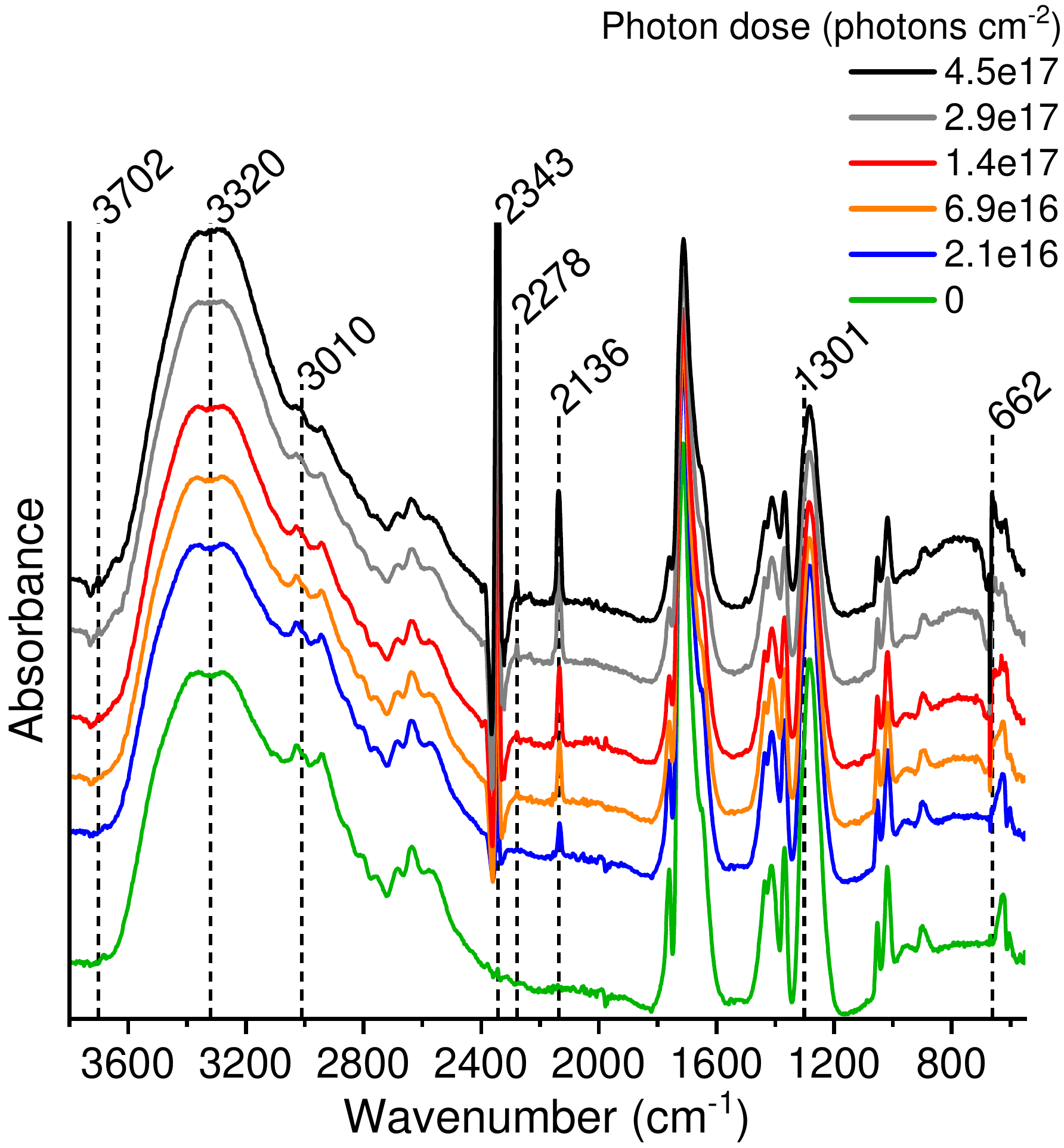}
  \caption{CH$_3$COOH:H$_2$O (Exp. \textbf{11})}
  \label{Fig.HAc_H2O_irradiacion}
\end{subfigure} %irradiación de HAc:H2O
\caption{Evolution of the IR spectra during the irradiation of three different CH$_3$COOH-bearing ice samples at 10 K. Vertical lines indicate the appearance of new vibrational bands, whose assignments are given in Table \ref{Table.IR_photoproducts}.}
\label{Fig.HAc_UV_irradiation}
\end{figure}

Fig. \ref{Fig.HAc_UV_irradiation} shows the IR spectra obtained in each period of UV irradiation of acetic acid ice samples at 10 K. These spectra show the gradual destruction of acetic acid, which is reflected in the decrease of its IR bands. The appearance of new IR bands related to formation of new species is observed, which are highlighted by vertical lines in the spectrum. Table \ref{Table.IR_photoproducts} details the features that emerged during the irradiation of the ices, as well as the photoproducts associated with each of them.\\

In the Xe ice matrix, UV irradiation is expected to dissociate CH$_3$COOH into CH$_3$CO$\cdot$, $\cdot$COOH, H$\cdot$, $\cdot$CH$_3$ and $\cdot$OH radicals \citep{Macoas2004}. CH$_4$ is detected at 3007 and 1300 cm$^{-1}$ and CH$_3$CH$_3$, at 2973 and 2880 cm$^{-1}$, probably formed by the recombination of the CH$_3$$\cdot$ radicals \citep[]{Carrascosa2020}. Decomposition of the COOH radical could lead to the formation of CO$_2$, detected by the IR features at 2338, 2274 ($^{13}$CO$_2$) and 660 cm$^{-1}$. Other photoproducts detected are CO (2134 cm$^{-1}$) and H$_2$O (broadband around 3321 cm$^{-1}$). According to \cite{Macoas2004}, CO and H$_2$O can be formed via different mechanisms from CH$_3$CO$\cdot$ and OH$\cdot$ radicals, although other pathways such as CO formation from UV irradiation of CO$_2$ could be relevant in our experiments.\\
Photoproducts detected in the Xe matrix are also found in pure CH$_3$COOH and CH$_3$COOH:H$_2$O ice samples (Exps. \textbf{5} and \textbf{11}). However, formation of hydrocarbon species such as CH$_4$ and CH$_3$CH$_3$ is hindered in the CH$_3$COOH:H$_2$O mixture (only CH$_4$ was observed with low abundance) probably due to the presence of O-H groups playing an active role in the chemical reaction network.\\

 %IR de HAc, HAc:Xe y HAc:H2O irradiados  

\begin{table*} %Bandas IR formadas tras UV de los hielos
\small
   %\centering
\caption[]{New IR band positions (in cm$^{-1}$) detected in the infrared spectra of the ice samples during UV irradiation at 10 K, and their identification.}
    \label{Table.IR_photoproducts}
    %\resizebox{8.5cm}{!} {
%\begin{center}
\begin{tabular}{lccccl}
\hline
\noalign{\smallskip}
Species & Vibration mode & CH$_3$COOH:Xe & CH$_3$COOH & CH$_3$COOH:H$_2$O & Reference \\
\noalign{\smallskip}
\hline
\noalign{\smallskip}
    CO$_2$&         comb.&           -&        3698&     3702&\cite{Bouilloud2015, Gerakines1995}\\
\noalign{\smallskip}
    H$_2$O&         O-H str.&           3321&       3318&      3320&\cite{Gerakines1995, HAGEN1981}\\
\noalign{\smallskip}
    CH$_4$&         C-H str.&           3007&       3008&      3009&\cite{RACHID2017, D'Hendecourt1986}\\
\noalign{\smallskip}
    CH$_3$CH$_3$&   CH$_3$ deg. str.&            2973&        2979&     -&\cite{Bennett_Kaiser2007, Gerakines1995}\\
\noalign{\smallskip}
    CH$_3$CH$_3$&   CH$_3$ sym. str.&           2880&        2886&     -&\cite{Bennett_Kaiser2007, Gerakines1995}\\
\noalign{\smallskip}
    CO$_2$&         C=O str.&           2338&        2340&     2343&\cite{Modica2010, Gerakines1995}\\
\noalign{\smallskip}
    $^{13}$CO$_2$&  C=O str.&           2274&         2276&    2278&\cite{Bouilloud2015, Modica2010}\\
\noalign{\smallskip}
    CO&             CO str.&           2134&         2134&    2136&\cite{RACHID2017, Gerakines1995}\\
\noalign{\smallskip}
    CH$_4$&         C-H bend.&           1300&        1302&     1301&\cite{Bouilloud2015, D'Hendecourt1986}\\
\noalign{\smallskip}
    CO$_2$&     O=C=O bend.&            660&         652&    662&\cite{Bouilloud2015, D'Hendecourt1986}\\
\noalign{\smallskip}
\hline
\end{tabular}\\
%\end{center}
%}
\begin{flushleft}
\textit{comb. = combination; str. = stretching; sym. = symmetric; bend. = bending; def. = deformation\\
*tentatively assigned.}\\
\end{flushleft}
%COMENTARIOS: \\
\end{table*} %Bandas IR formadas tras UV de los hielos

\subsubsection{Photodestruction and photoformation cross-sections}
\label{crosssections}

\begin{figure*} %Densidad de columna HAc, CO, CO2
    \centering
    \includegraphics[width=1\textwidth]{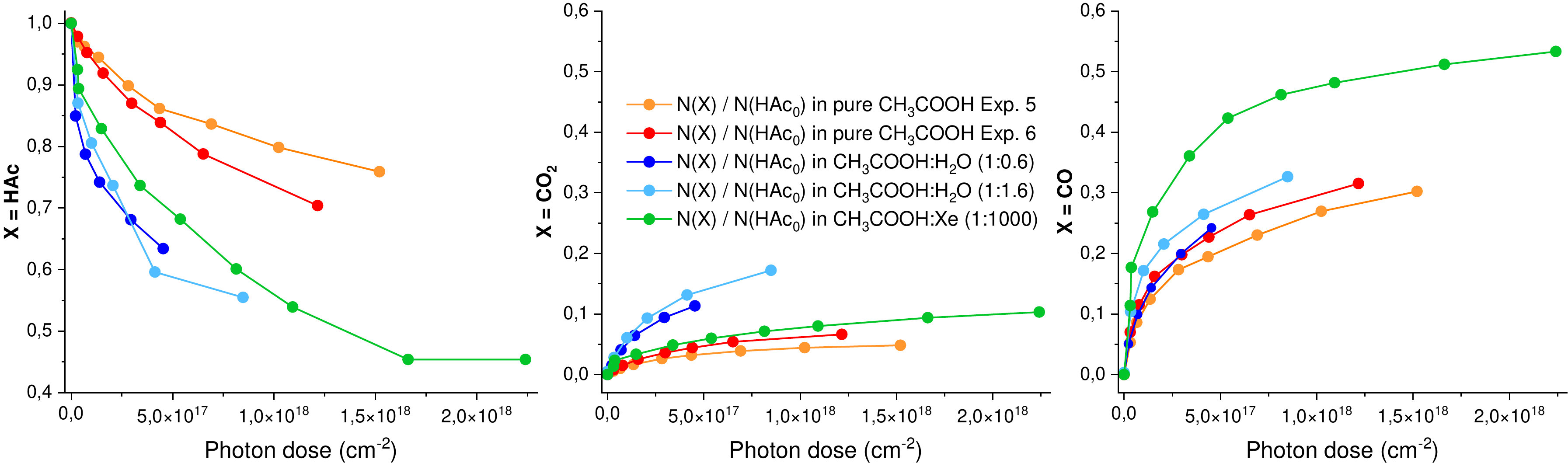}
    \caption{Relative column density of acetic acid (HAc, left) and the formation of CO$_2$ (middle) and CO (right) as a function of the incident photon dose, for the three ice samples studied in this work: pure CH$_3$COOH (Exp. \textbf{5} and \textbf{6}), CH$_3$COOH:H$_2$O (Exp. \textbf{11} and \textbf{12}) and CH$_3$COOH:Xe (Exp. \textbf{9}).} 
    \label{Fig.column_density}
\end{figure*}

Fig. \ref{Fig.column_density} shows the evolution of the abundances of CH$_3$COOH, CO$_2$ and CO in each ice sample. Their destruction and formation cross-sections, calculated using the Eqs. \ref{Eq.photodestruction.} and \ref{Eq.photoformation}, are given in Table \ref{Table.cross-sections}. Photodestruction cross-section of acetic acid is higher in CH$_3$COOH:Xe and CH$_3$COOH:H$_2$O ice samples. This is due to the effect of the Xe matrix and water molecules, respectively, which slow down the diffusion of radicals in the ice, trapping them in the ice matrix and reducing CH$_3$COOH recombination reactions. This trapping of radicals could explain why dissociated CH$_3$COOH is not converted into more complex molecules at 10 K \citep[]{Oberg2010}, even in pure acetic acid ice samples. On the other hand, $\sigma_{\text{des(CH$_3$COOH)}}$ varies significantly between CH$_3$COOH:H$_2$O ice samples. The rate of photodestruction of acetic acid molecules decreases as the amount of H$_2$O present in the mixture increases (Experiments \textbf{11}, \textbf{12} and \textbf{13}). A likely explanation is that, for the same UV dose, acetic acid molecules absorb less UV photons, because the increasing H$_2$O abundance absorbs a fraction of the UV photons.\\

A relatively large amount of CO$_2$ and CO is produced in the Xe matrix, due to the isolation effect of the photoproducts by the Xe atoms, which inhibits backwards reactions to reform the parent molecules. CH$_3$COOH:H$_2$O ice samples show a high rate of CO$_2$ formation, probably because the presence of H$_2$O favours the cleavage of the COOH radical, generating CO$_2$, H$_2$, and H$_2$O. CO$_2$ is further processed by UV photons, resulting in higher CO formation in CH$_3$COOH:H$_2$O compared to pure CH$_3$COOH ices, as it can be seen in Fig. \ref{Fig.column_density} (right panel). The CO formation cross-section was calculated from the destruction of acetic acid itself, without considering the contribution of water. Although the presence of H$_2$O facilitates CO formation, the CO formation efficiency does not increase as the water content of the mixture increases, because water molecules also reduce the number of UV photons absorbed by CH$_3$COOH molecules, thus limiting their photolysis and subsequent formation of CO$_2$ and CO.\\

 %Densidad de columna HAc, CO, CO2

\begin{table*} %Destruction & Formation cross-section
\small
   %\centering
\caption[]{Initial destruction and formation cross-sections (in cm$^{2}$) for each ice sample at 10 K. In the case of pure acetic acid ice, the values correspond to the average of three experiments (Exp. \textbf{5}, \textbf{6} and \textbf{7}).}
    \label{Table.cross-sections}
    %\resizebox{15cm}{!} {
%\begin{center}
\begin{tabular}{lccccc}
\hline
\noalign{\smallskip}
 & CH$_3$COOH  & CH$_3$COOH:H$_2$O (1:0.6) & CH$_3$COOH:H$_2$O (1:1.6) & CH$_3$COOH:H$_2$O (1:2.5) & CH$_3$COOH:Xe \\
\noalign{\smallskip}
\hline
\noalign{\smallskip}
$\sigma_{\text{des}}$ CH$_3$COOH & ($6.7\pm3.1)\times10^{-19}$ & $7.6\times10^{-18}$ & $4.1\times10^{-18}$ & $3.9\times10^{-18}$ & $2.5\times10^{-18}$\\
\noalign{\smallskip}
{$\sigma_{\text{form}}$} CO$_2$ & ($6.8\pm1.6)\times10^{-22}$ & $1.3\times10^{-21}$ & $1.5\times10^{-21}$ & $1.0\times10^{-21}$ & $1.5\times10^{-21}$\\
\noalign{\smallskip}
{$\sigma_{\text{form}}$} CO & ($6.3\pm1.1)\times10^{-21}$ & $4.3\times10^{-21}$ & $5.4\times10^{-21}$ & $4.9\times10^{-21}$ & $1.3\times10^{-20}$\\
\noalign{\smallskip}
\hline
\end{tabular}\\
%}
\end{table*} %Destruction & Formation cross-section 

\subsubsection{CO photodesorption during UV irradiation of CH$_3$COOH ice samples}
\label{CO_photodesorption}

\begin{figure*} %Fotodesorción CO
  \centering 
\begin{subfigure}{.49\textwidth} %QMS Fotodesorción CO
  \centering 
  \includegraphics[width=\linewidth]{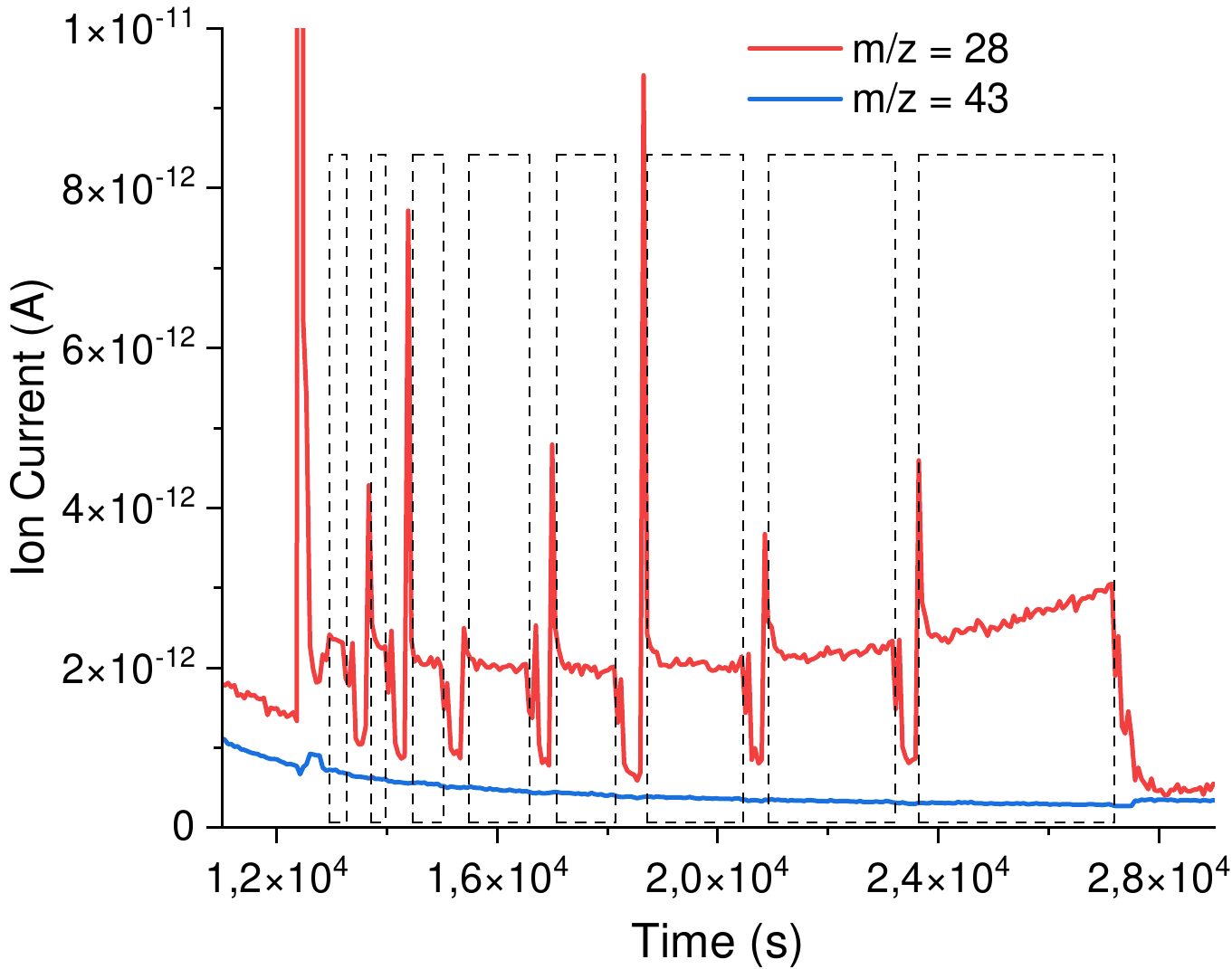}
  \label{Fig.QMS_CO_photodesorption}
\end{subfigure} %QMS Fotodesorción CO
\begin{subfigure}{.49\textwidth} %Fotodesorción CO IR
    \centering
    \includegraphics[width=\textwidth]{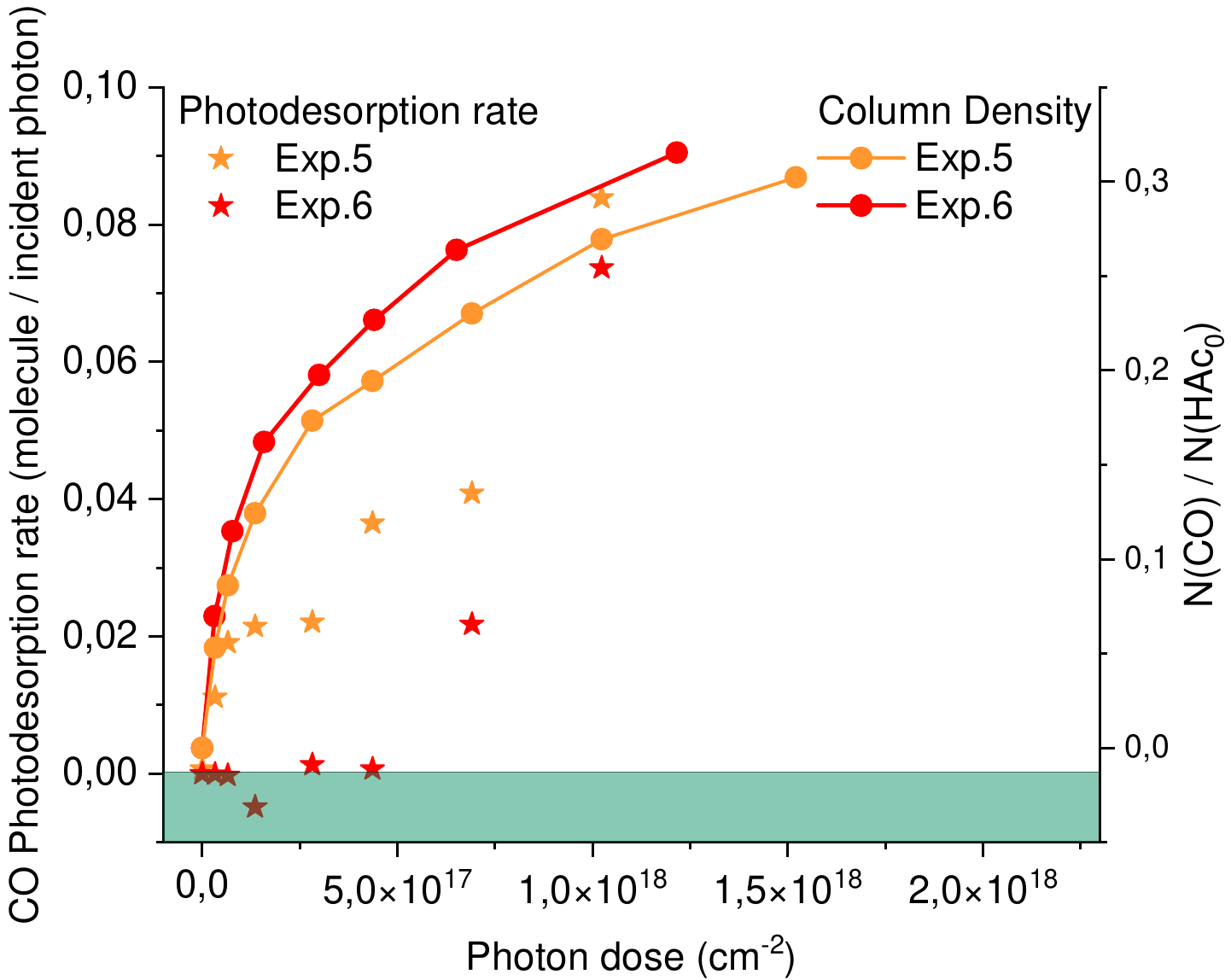}
    \label{Fig.photodesorption}
\end{subfigure} %Fotodesorción CO IR
\caption{Left panel: increase of the QMS ion current intensities during the various irradiation intervals of CH$_3$COOH ice samples (Exp. \textbf{5}). The signal $\frac{m}{z}$ = 28 corresponds to CO molecules and the signal $\frac{m}{z}$ = 43 to a fragment which is not expected to be photodesorbed. Right panel: CO photodesorption (left axis) and CO relative column density (right axis) in pure CH$_3$COOH ice samples (Exp. \textbf{5} and Exp. \textbf{6}) as a function of the incident UV photon dose. }
\label{Fig.CO_photodesorption}
\end{figure*}

CO column density increases readily during the first stages of pure CH$_3$COOH ices UV-irradiation (Exps. \textbf{5} and \textbf{6}, Fig \ref{Fig.column_density}). For large irradiation doses, CO column density stabilises due to the lower availability of acetic acid molecules to form CO. Thus, the initial increase in CO column density results in increased CO availability at the ice surface. The absorption of a photon by CO molecules at the surface induces their desorption to the gas phase, as it can be seen in the left panel of Fig. \ref{Fig.CO_photodesorption}. In addition, the electronic excitation energy is redistributed to neighboring molecules when CO returns to the ground state, what may result in the photodesorption of several molecules, as observed in the increase of CO photodesorption along the irradiation steps (Fig. \ref{Fig.CO_photodesorption}, right panel).\\

 %Fotodesorción CO 

%%%%%%%%%%%%%%%%%%%%%%%%%%%%%%%%%%%SECTION ÁCIDO OXÁLICO%%%%%%%%%%%%%%%%%%%%%%%%%%%%%%%%%%%%%%%
\subsection{Synthesis of organic molecules during warm-up of UV irradiated CH$_3$COOH ice samples.}
\label{sect.warmup}

\begin{figure} %IR de calentamiento de HAc UV
    %\centering
    \includegraphics[width=0.5\textwidth]{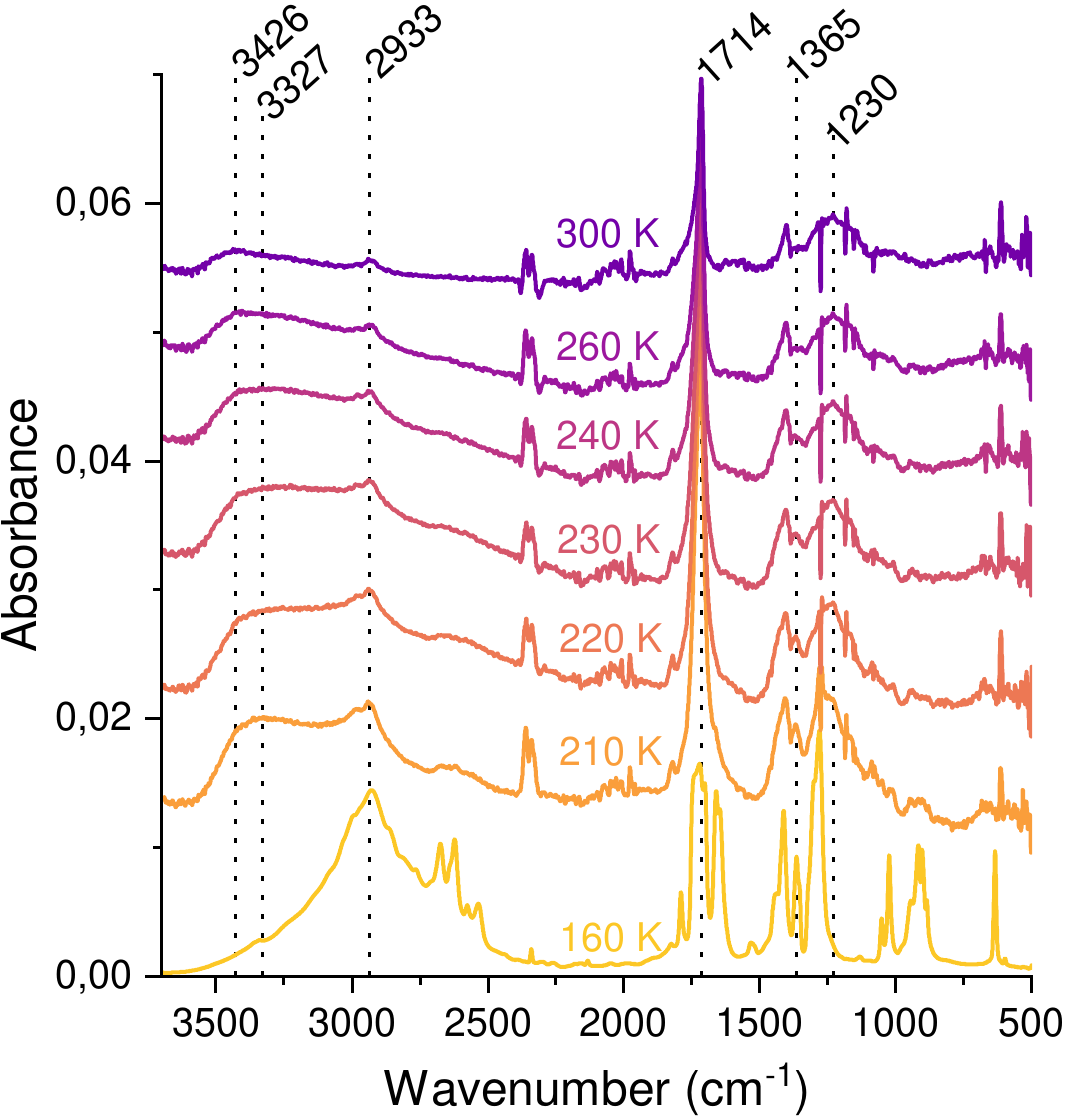}
    \caption{Evolution of the IR spectrum during warm-up of UV-irradiated CH$_3$COOH ice sample (Exp. \textbf{8}). The vertical lines highlight the features that undergo changes.}
    \label{Fig.warmupHAc}
\end{figure}

\begin{figure} %IR de calentamiento de HAc:H2O UV
    \centering
    \includegraphics[width=0.5\textwidth]{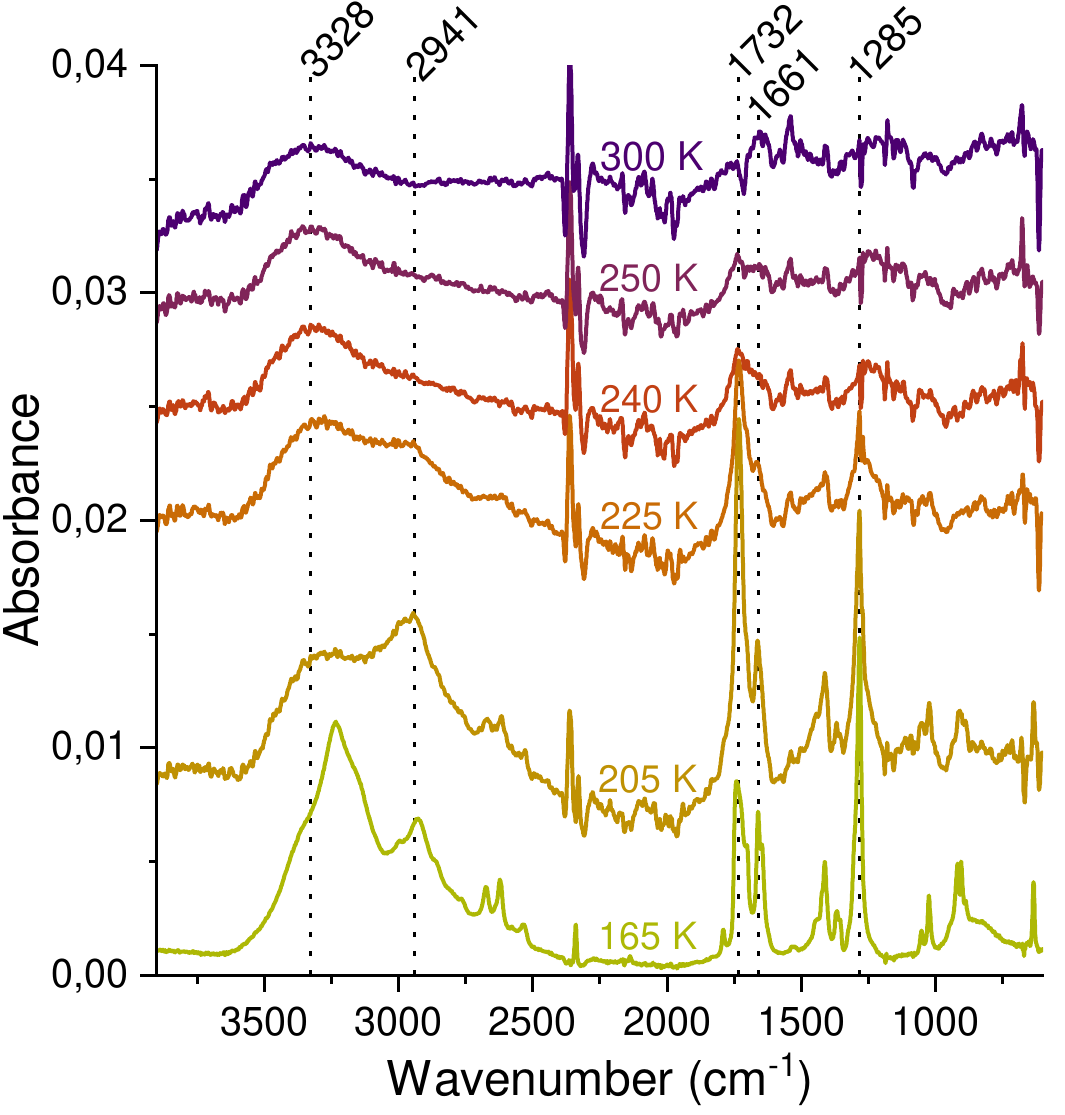}
    \caption{Infrared spectra of the CH$_3$COOH:H$_2$O ice mixture (Exp. \textbf{11}) at different temperatures during warm-up. The vertical lines highlight the features that undergo changes.}
    \label{Fig.warmupIRhach2o}
\end{figure}

Warm-up of the UV-irradiated ice samples causes changes in their composition. Small volatile molecules, as CO, CO$_2$, CH$_4$ and CH$_3$CH$_3$, formed during irradiation, are ejected into the gas phase at relatively low temperatures (T<100 K). In addition, heating of the ice increases the mobility of the radicals formed during irradiation. Then, these radicals react to form more complex species. At 205-210 K, above CH$_3$COOH thermal desorption (160-190 K), the IR spectra of the ice samples still display features corresponding to acetic acid. This is observed in both, pure acetic acid and binary mixtures with water (Fig. \ref{Fig.warmupHAc} and \ref{Fig.warmupIRhach2o}). The reason for this is that complex molecules during warm-up form a matrix in the ice that establishes intermolecular bonds with the acetic acid molecules, preventing CH$_3$COOH desorption.\\

 %IR de calentamiento de HAc UV

 %IR de calentamiento de HAc:H2O UV

\subsubsection{Warm-up of irradiated pure CH$_3$COOH ices}
\label{warmupHAc}

As temperature increases, a gradual decrease in the intensity of the IR bands is observed until room temperature for pure CH$_3$COOH ice samples, forming a residue (Fig. \ref{Fig.warmupHAc}). During this process, the positions of the C=O (1714 cm$^{-1}$) and CH$_3$ (2933 cm$^{-1}$) stretching modes remain constant, suggesting that the residue contains acetic acid-derived species. However, the hydroxyl groups in these species are in different molecular environments to those of acetic acid, evidenced by the shift of the OH vibration from 3327 to 3426 cm$^{-1}$. In addition, the C-O band of acetic acid (1280 cm$^{-1}$) disappears during warm-up, leaving a new band observable at 1230 cm$^{-1}$, which appears at 210 K. This new band can be attributed to the C-O vibration of carboxylic acid groups. Furthermore, the decrease in the intensity of the {$\delta$}CH$_3$ band (1365 cm$^{-1}$) indicates that the new species formed have longer chains, since the CH$_3$ groups are present at the end of the molecule.\\

\subsubsection{Warm-up of irradiated CH$_3$COOH:H$_2$O ices}
\label{warmupHAcH2O}

\begin{figure} %QMS desorción tardía HAc en HAc:H2O UV
    \centering
    \includegraphics[width=0.3\textwidth]{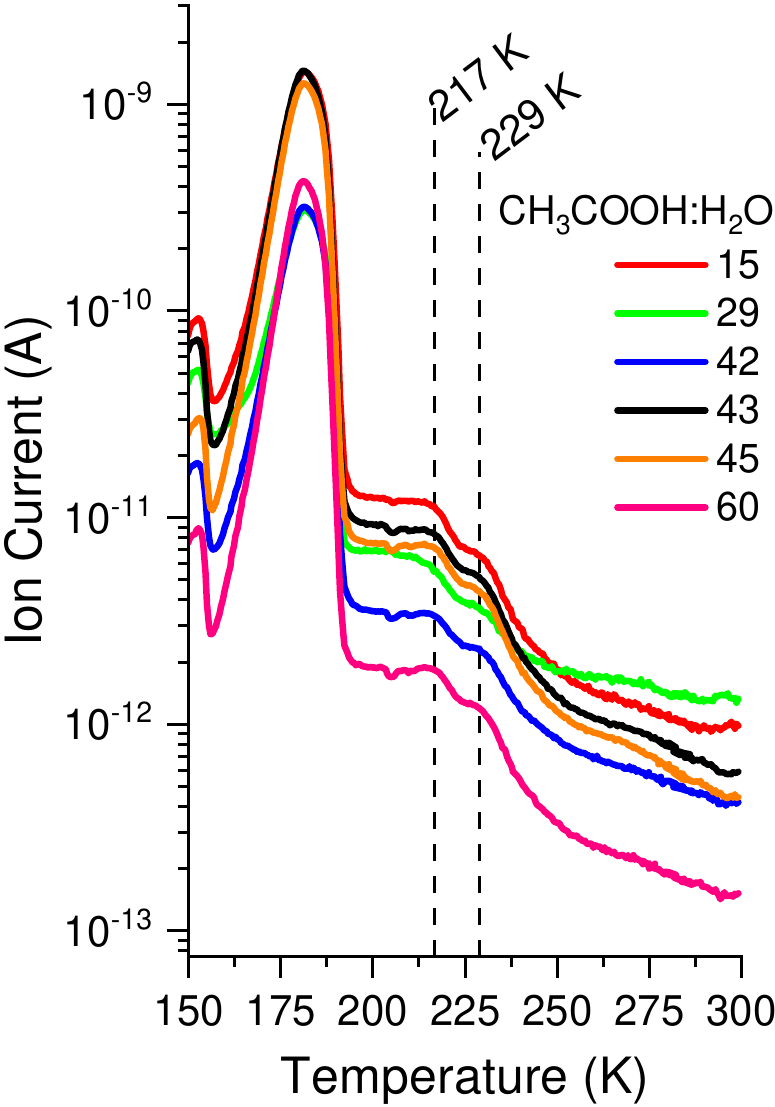}
    \caption{TPD of the CH$_3$COOH:H$_2$O (1:0.6) ice sample (Exp. \textbf{11}). The vertical lines at temperatures 217 and 229 K indicate the late desorption of acetic acid trapped in the matrix of complex molecules that have formed in the ice after photon and thermal processing. $\frac{m}{z}$= 20 is not expected to correspond to any molecule and is used as a reference.}
    \label{Fig.warmupQMShach2o}
\end{figure}

In the case of the CH$_3$COOH:H$_2$O ice samples, significant changes in the IR spectrum are observed between 205 and 240 K (see Fig. \ref{Fig.warmupIRhach2o}). In this temperature range, IR band of CH$_3$ (2941 cm$^{-1}$) vibrations disappear and the C=O at 1732 cm$^{-1}$ and C-O at 1285 cm$^{-1}$ experience a pronounced decrease in intensity. At the same time, data collected by QMS, shown in Fig. \ref{Fig.warmupQMShach2o}, indicate that the ion current measured for $\frac{m}{z}$ = 60, 45, 43, 42, 29 and 15 fragments, associated to acetic acid molecules, exhibit desorption peaks at 217 and 229 K. Altogether, both IR and QMS analysis indicate that the remaining acetic acid molecules desorb between 205-240 K.\\

 %QMS desorción tardía HAc en HAc:H2O UV

\begin{figure*} % Ácido Oxálico
\begin{subfigure}{.20\textwidth} % QMS Oxálico HAc:H2O
  \centering 
  \includegraphics[width=\linewidth]{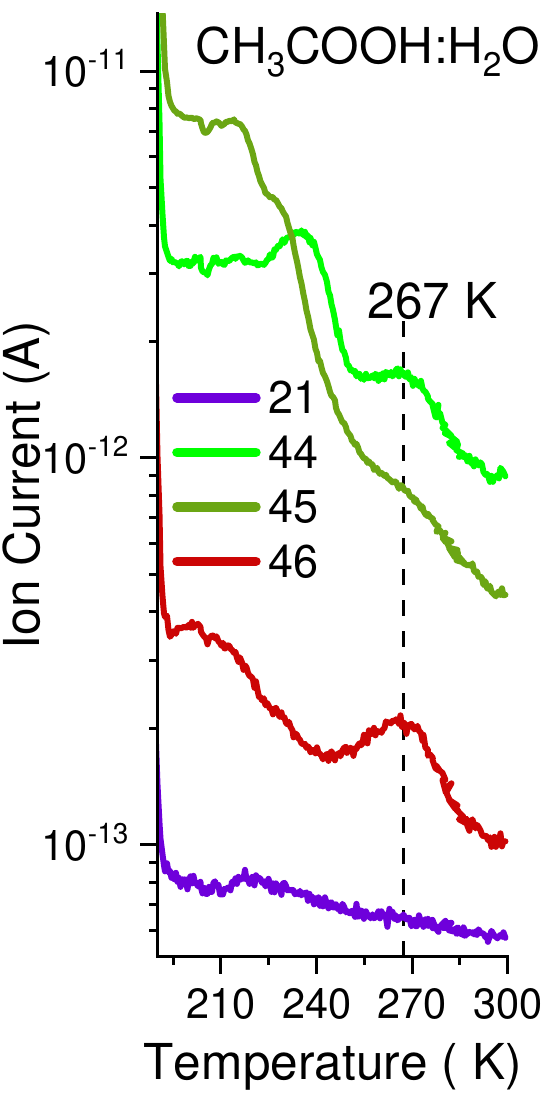}
\end{subfigure} % QMS Oxálico HAc:H2O 
\begin{subfigure}{.45\textwidth} %Oxálico NIST
  \includegraphics[width=\linewidth]{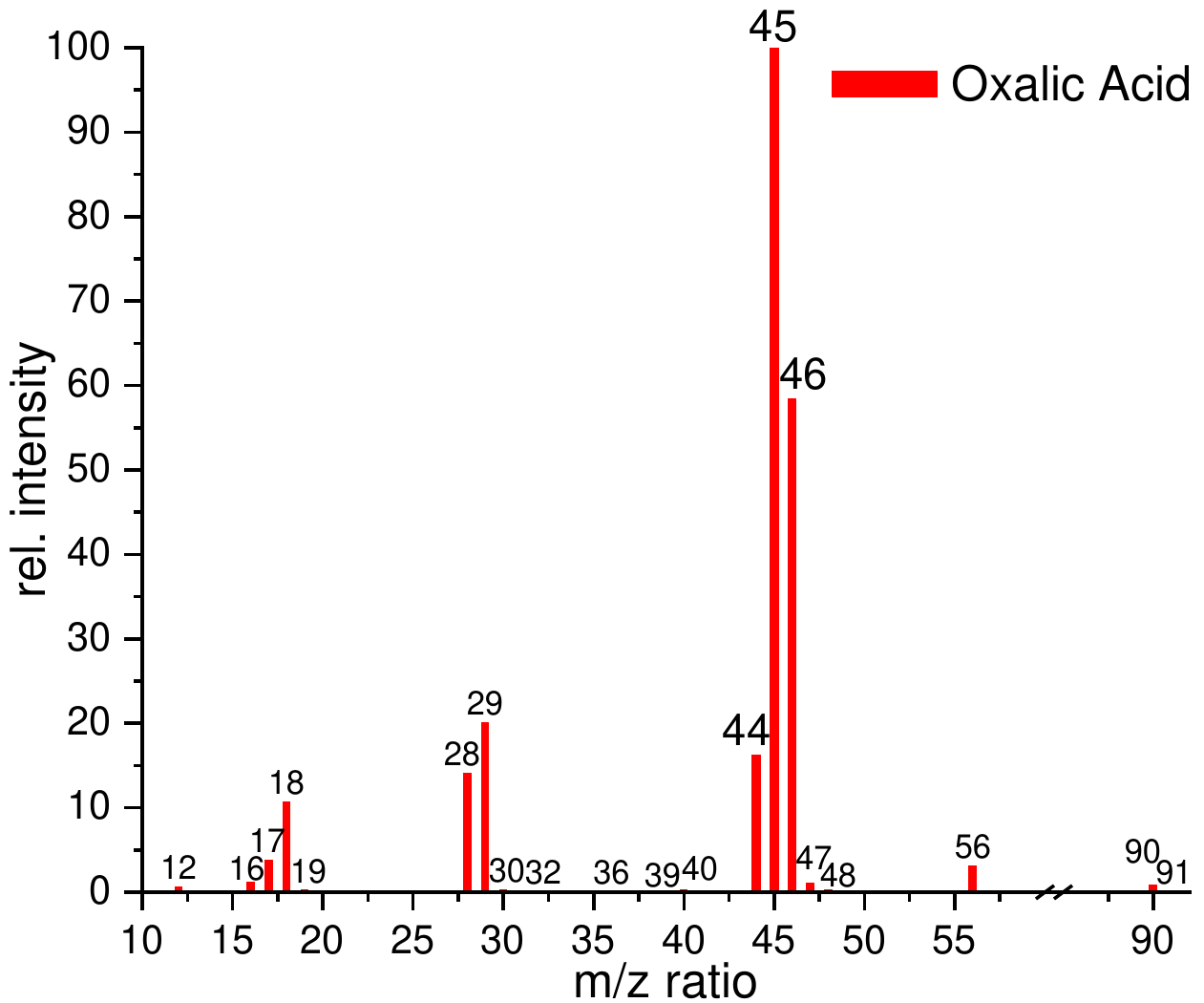}
\end{subfigure} %Oxálico NIST
\begin{subfigure}{.34\textwidth} % Síntesis Ácido Oxálico
  \includegraphics[width=\linewidth]{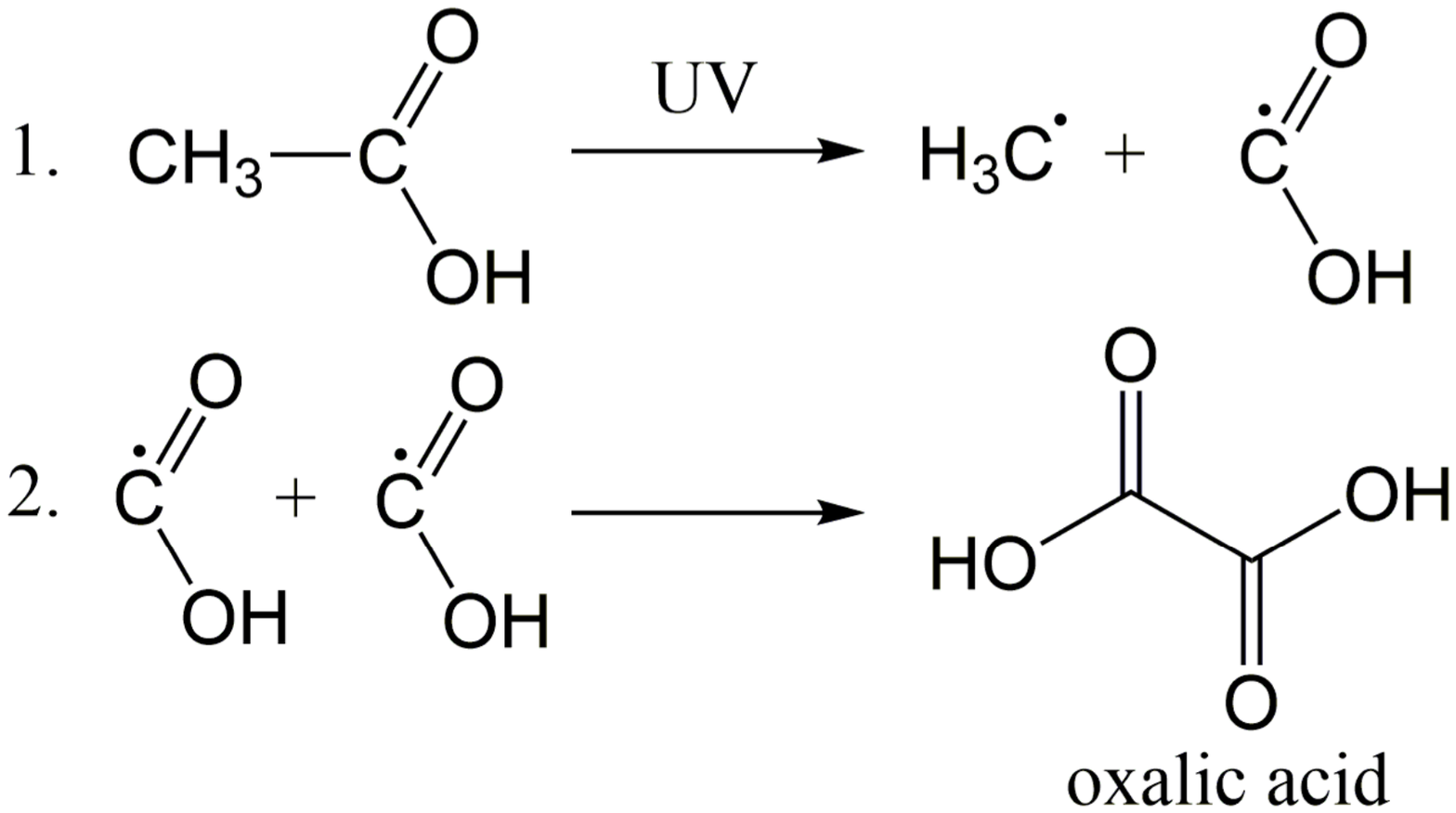}
\end{subfigure} % Síntesis Ácido Oxálico
\caption{Tentative detection of oxalic acid (HOOC-COOH) during warm-up of an UV-irradiated CH$_3$COOH:H$_2$O ice sample (Exp. \textbf{11}). Left panel: QMS signal for $\frac{m}{z}$= 44, 45 and 46. These fragments are representative of the oxalic acid, while $\frac{m}{z}$= 21 is not expected to correspond to any molecule and is used as a reference. Centre panel: mass spectrum of oxalic acid taken from NIST database. Right panel: scheme to form oxalic acid via radical-radical reaction from photon and thermal processing of acetic acid-containing ices. The formation temperature of oxalic acid could not be determined, since oxalic acid was only detected from QMS data in the gas phase during thermal desorption.}
  \label{Fig.oxalico}
\end{figure*} %Ácido Oxálico

The C=O stretching band of carboxylic acids (1732 cm$^{-1}$) disappears between 250 and 280 K in CH$_3$COOH:H$_2$O ice samples. Data recorded by QMS in this temperature range are presented in the left panel of Fig. \ref{Fig.oxalico}. This figure shows that the ion current corresponding to $\frac{m}{z}$ = 46, 45 and 44 exhibit a maximum peak desorption at 267 K. These $\frac{m}{z}$ signals can be associated with COO$^{+}$ and COOH$^{+}$ fragments, suggesting that the species responsible for this thermal desorption consists of carboxyl groups. Therefore, a possible candidate could be oxalic acid (HOOC-COOH), whose infrared spectrum (\cite{Muthuselvi2016}) is in good agreement with the disappearance C=O stretching bands in this temperature range and mass spectrum is consistent with the fragments observed in our QMS data (Fig. \ref{Fig.oxalico}, middle panel). \cite{Warren1959} reported the formation of oxalic acid after proton bombardment of acetic acid in oxygenated aqueous solutions. As reported by \cite{Huiyan2008}, desorption of oxalic acid takes place around 241 K in UHV experiments. In our experiments, desorption of oxalic acid may be delayed due to the presence of other complex molecules remaining in the irradiated ice. Oxalic acid could have formed in the ice from direct recombination of $\cdot$COOH radicals (Fig. \ref{Fig.oxalico}, right panel), which acquire mobility with increasing temperature. After desorption of oxalic acid, warm-up to 300 K does not cause significant changes in the IR spectrum. A residue is observed at room temperature containing mainly a broad band corresponding to the O-H vibration at 3328 cm$^{-1}$.\\

%%%%%%%%%%%%%%%%%%%%%%%%%%%%%%%%%%%SECCIÓN CONCLUSIONES%%%%%%%%%%%%%%%%%%%%%%%%%%%%%%%%%%%%%%%%%%%%%
\section{Conclusions}
\label{sect.conclusions}
Thermal processing of CH$_3$COOH, CH$_3$COOH:H$_2$O and CH$_3$COOH:Xe ice samples allowed to characterise the structural changes in the IR spectra and TPD curves (Sect. \ref{sect.enviroments}):\\

\begin{itemize}

\item	Pure acetic acid ice deposited at 10 K presents an arrangement in the form of cyclic dimers and amorphous monomers trapped in the bulk. At 120 K, restructuring of the monomers to cyclic dimers takes place. Phase transformation to a crystalline organisation in the form of chain polymers proceeds above 160 K. Acetic acid molecules then sublimate around 189 K.\\

\item	In CH$_3$COOH:H$_2$O ice mixture (1:1), interaction of the O-H groups causes a broadening of the IR bands. Crystallisation of both species occurs simultaneously at 149 K, while no evidence was found for the formation of a molecular complex.\\

UV irradiation of the ices studied in this paper at 10 K leads to the gradual destruction of acetic acid and mainly to the formation of CO, CO$_2$, CH$_4$, CH$_3$CH$_3$ and H$_2$O (Sect. \ref{sect.UV_irradiation}):\\

\item	The presence of O-H in CH$_3$COOH:H$_2$O ice samples prevents the formation of hydrocarbon molecules such as ethane (CH$_3$CH$_3$).\\

\item	The presence of water molecules and Xe in the ice forbids recombination of the radicals formed after UV irradiation. This results, on one hand, in higher rates of formation of small molecules such as CO$_2$ and consequently CO. On the other hand, higher rates of photodestruction of CH$_3$COOH were obtained.\\

\item	As the proportion of water in the ice mixture increases, it inhibits the absorption of UV photons by acetic acid molecules. Therefore, a lower number of radicals are formed, thus reducing the rates of photodestruction of CH$_3$COOH and the formation of small molecules such as CO$_2$ and CO.\\

\item	In pure CH$_3$COOH ices, accumulation of CO molecules on the ice surface leads to photodesorption of these molecules to the gas phase.\\

\item For the first time, the VUV-absorption cross-section of CH$_3$COOH in the solid phase has been measured, with an average value of (6.0 $\pm$ 0.4) $\times$ 10$^{-18}$ cm$^{2}$. Our estimated values of the VUV absorption cross-section within the spectral range of the H$_2$ Lyman system are comparable to those reported by \cite{Leach2006} for acetic acid in the gas phase.\\

Warm-up of the UV-irradiated ice samples in this work leads to the formation of more complex species (Sect. \ref{sect.warmup}):\\

\item The complex molecules formed during warm-up of UV-processed pure CH$_3$COOH ice samples form a matrix that establishes intermolecular bonds with acetic acid. This explains the desorption temperature of 210 K, i.e. higher than the one observed in pure acetic acid ice, at 190 K.\\

\item Warm-up of the irradiated CH$_3$COOH ice samples to room temperature shows the formation of a residue composed of long aliphatic chains with hydroxyl (O-H) groups displaying different IR features from those of acetic acid and CO groups typical of carboxylic acids.\\

\item The formation of oxalic acid (HOOC-COOH) in ice samples CH$_3$COOH:H$_2$O is proposed. TPD experiments show its desorption at 267 K through fragments of mass m/z = 46, 45 and 44. Furthermore, the analysis of the IR spectra showed the disappearance of a carbonyl group in this temperature range, what may support the thermal desorption of oxalic acid.\\

\item Warm-up of irradiated CH$_3$COOH:H$_2$O ice mixtures to room temperature leaves a residue that displays mainly a broad IR band corresponding to the O-H stretching mode. The most important effect of the presence of water in the ice is the evolution from a chemistry based on aliphatic chains towards products with a more polar character. This transformation is evidenced by comparing the IR spectra of the residue produced in CH$_3$COOH ices (Fig. \ref{Fig.warmupHAc}), where the C-H groups remain, with those of CH$_3$COOH:H$_2$O ices (Fig. \ref{Fig.warmupIRhach2o}), where the C-H groups have disappeared and the O-H groups predominate.\\
\end{itemize}

\section{Astrophysical implications}
\label{sect.astrophysical_implications}

Acetic acid has been detected in the gas phase toward Sgr B2 \citep{Mehringer1997}. More recently, CH$_3$COOH was also observed in the warm inner regions of the envelopes surrounding protostars in the binary system IRAS 16293–2422 \citep{manigand2020}. It displays a compact emission toward both protostars, suggesting that acetic acid is desorbed from ice mantles near the protostars, where temperatures enable its desorption near 187 K in our laboratory experiments, or about 116 K in space (calculated using the relationship described in \cite{Martin-Domenech2014}).\\         

Ongoing observations of ice mantles with JWST might serve to detect or set upper limits on the relative abundance of acetic acid. We found that the most intense band of acetic acid diluted in water is located at 5.85 $\mu$m (1709 cm$^{-1}$), which corresponds to the C=O stretching. The band falls in a relatively clean spectral region of the infrared spectrum of HH 46 IRS 1 (IRAS 08242-5050), a very young Class I protostellar object observed by Spitzer Space Telescope \citep[]{Noriega2004}. However, it should be noted that a number of molecules of considerable abundance have important bands in the same wavenumber range, such as H$_2$O, NH$_3$, HCOOH, and H$_2$CO. Acetic acid could be more easily detected in ice mantles exposed to relatively high temperature (around 100 K), as most of these species would have already desorbed.\\ %(ver espectros en papers de Schutte, Dartois y Boogert). %Mirar posicion de banda equivalente a 7.24 $\mu$m attributed to CH deformation mode of formic acid (HCOOH) y si se forma una cerca de 7.41 $\mu$m band compatible with the formate ion (HCOO$^-$) cuando se irradia acetic acid, no veo que la menciones en el texto, tal vez porque no la ves. (No se ve) 

%Decir que productos esperas encontrar en lines of sight donde el acetico es irradiado cuando se encuentra puro o con agua. Tal vez se pueda establecer una relacion con las observaciones en el gas de esas especies. Comentar sus temperaturas de desorcion empezando por la del acetico. Para ello hay que dar las que salen en la TPD y luego convertidas al caso espacial (como 1/3 mas bajas), ver Martin-Domenech 2014 para la conversion.
%Nota posterior 1: Tal vez se pueda decir algo sobre las especies que acompañan al acetico en base a nuestros experimentos, pero en base al paper que hemos mirado no creo que se pueda comparar con las observaciones porque intervienen mas factores: resiliencia en fase gaseosa debida a reacciones en el gas o diferente tasa de disociacion en el gas debida a la radiacion. 
%Nota posterior 2: Methoxymethanol (CH3OCH2OH) corresponds to the same category as acetic acid (Manigand et al. 2020), it desorbs within a temperature range similar to acetid acid,    https://iopscience.iop.org/article/10.3847/1538-4357/ab2989 but it displays extended emission. Why? May be because it is less efficiently destroyed in the gas phase? I cannot think why this would be the case.

\begin{table} %VUV-absorption cross section
\centering
\caption{Penetration depth of photons in pure acetic acid ice deposited at 10 K, expressed in column density capable of absorbing 95 and 99\% of the emitted photon flux.}
\label{table.VUVabsorption}
\begin{tabular}{|l|c|c|c|c|}
\hline
& \multicolumn{2}{c|}{95\% Photon Absorption} & \multicolumn{2}{c|}{99\% Photon Absorption} \\
Species & Avg. & Max. & Avg. & Max. \\
 & \multicolumn{2}{c|}{(×$10^{17}$ molecule cm$^{-2}$)} & \multicolumn{2}{c|}{(×$10^{17}$ molecule cm$^{-2}$)} \\ 
\hline
 CH$_3$COOH & 5.0 & 2.6 & 10.2 & 4.0\\
\hline
\end{tabular}
\begin{flushleft}
\textit{Avg. Corresponds to the average value of the cross-section in the 130-180 nm range and Max. corresponds to the maximum value of the cross-section in the same wavelength range.}\\
\end{flushleft}
\end{table} %VUV-absorption cross section

Penetration depth of photons in acetic acid ice was calculated from Eq. \ref{Eq.VUV_absorption.} and summarized in Table \ref{table.VUVabsorption}. VUV-absorption cross-sections of acetic acid ice can be used as input to calculate the absorption of multilayer ices, although they can not be extrapolated directly to ice mixtures. Instead, they could be valuable in models that simulate the UV processing of ice mantles present in cold astrophysical environments, such as the interior of a dense cloud and circumstellar regions.\\

Carboxylic acids are among the species less commonly detected in the interstellar medium. They are widely distributed in nature and are considered to be precursors of many relevant prebiotic molecules, such as amino acids and lipids \citep{SanzNovo2023}. We have reported the tentative formation of oxalic acid (HOOC-COOH) upon irradiation and warm-up of a CH$_3$COOH:H$_2$O ice sample. It was formed during desoprtion at 267 K, but its presence could not be confirmed in the IR, probably due to its low sensitivity compared to QMS. Oxalic acid is believed to have played an important role in abiotic surface-catalysed fatty acid synthesis \citep{Cohen2023}, as well as performing important functions in tropospheric chemistry \citep{Huiyan2008}. This species has not yet been detected in the interstellar medium. The conformational diagram of oxalic acid collected by \cite{Godfrey2000} shows six conformational isomers. The most abundant isomers (comprising 82\% of the total oxalic acid) will not be easily detected because they exhibit an extremely low dipole moment $(\mu_{\text{tot}}$ = 0.6 D for the cis-trans-cis and $\mu_{\text{tot}}$ = 0 D for the trans-trans-trans isomer). Finally, our experimental study may serve to understand the chemical connection of acetic acid with other species detected toward the same line of sight. \\

\section{Acknowledgements}

This research has been funded by project PID2020-118974GB-
C21 by the Spanish Ministry of Science and Innovation. C.B. was supported by an INTA grant. B.E. acknowledges support by grant PTA2020-
018247-I by the Spanish Ministry of Science and Innovation/State
Agency of Research MCIN/AEI. R.M has received funding from “la Caixa” Foundation, under agreement LCF/BQ/PI22/11910030. 

\section{Data availability}
The data underlying this article will be shared on reasonable request to the corresponding authors.\\

%%%%%%%%%%%%%%%%%%%%%%%%%%%%%%%% REFERENCES %%%%%%%%%%%%%%%%%%%%%%%%%%%%%%%%%%%%%
\bibliographystyle{mnras}
\bibliography{bibliography} % if your bibtex file is called example.bib

\begin{thebibliography}{}
\makeatletter
\relax
\def\mn@urlcharsother{\let\do\@makeother \do\$\do\&\do\#\do\^\do\_\do\%\do\~}
\def\mn@doi{\begingroup\mn@urlcharsother \@ifnextchar [ {\mn@doi@}
  {\mn@doi@[]}}
\def\mn@doi@[#1]#2{\def\@tempa{#1}\ifx\@tempa\@empty \href
  {http://dx.doi.org/#2} {doi:#2}\else \href {http://dx.doi.org/#2} {#1}\fi
  \endgroup}
\def\mn@eprint#1#2{\mn@eprint@#1:#2::\@nil}
\def\mn@eprint@arXiv#1{\href {http://arxiv.org/abs/#1} {{\tt arXiv:#1}}}
\def\mn@eprint@dblp#1{\href {http://dblp.uni-trier.de/rec/bibtex/#1.xml}
  {dblp:#1}}
\def\mn@eprint@#1:#2:#3:#4\@nil{\def\@tempa {#1}\def\@tempb {#2}\def\@tempc
  {#3}\ifx \@tempc \@empty \let \@tempc \@tempb \let \@tempb \@tempa \fi \ifx
  \@tempb \@empty \def\@tempb {arXiv}\fi \@ifundefined
  {mn@eprint@\@tempb}{\@tempb:\@tempc}{\expandafter \expandafter \csname
  mn@eprint@\@tempb\endcsname \expandafter{\@tempc}}}

\bibitem[\protect\citeauthoryear{Ahmad, Shivani, Misra  \& Tandon}{Ahmad
  et~al.}{2020}]{Ahmad2020}
Ahmad A.,  Shivani Misra A.,   Tandon P.,  2020, \mn@doi [Research in Astronomy
  and Astrophysics] {10.1088/1674-4527/20/1/14}, 20, 014

\bibitem[\protect\citeauthoryear{Altwegg et~al.,}{Altwegg
  et~al.}{2017}]{Altwegg2017}
Altwegg K.,  et~al., 2017, \mn@doi [Monthly Notices of the Royal Astronomical
  Society] {10.1093/mnras/stx1415}, 469, S130

\bibitem[\protect\citeauthoryear{Andersen}{Andersen}{1980}]{Andersen1980}
Andersen H.~C.,  1980, J. Chem. Phys., 71, 2384

\bibitem[\protect\citeauthoryear{Bahr, Borodin, Höfft, Kempter, Allouche,
  Borget  \& Chiavassa}{Bahr et~al.}{2006}]{Bahr&Borodin2006}
Bahr S.,  Borodin A.,  Höfft O.,  Kempter V.,  Allouche A.,  Borget F.,
  Chiavassa T.,  2006, \mn@doi [The Journal of Physical Chemistry B]
  {10.1021/jp055980u}, 110, 8649

\bibitem[\protect\citeauthoryear{Bennett \& Kaiser}{Bennett \&
  Kaiser}{2007}]{Bennett_Kaiser2007}
Bennett C.~J.,  Kaiser R.~I.,  2007, The Astrophysical Journal, 660, 1289

\bibitem[\protect\citeauthoryear{Bergantini, Zhu  \& Kaiser}{Bergantini
  et~al.}{2018}]{Bergantini2018}
Bergantini A.,  Zhu C.,   Kaiser R.~I.,  2018, \mn@doi [The Astrophysical
  Journal] {10.3847/1538-4357/aacf93}, 862, 140

\bibitem[\protect\citeauthoryear{Bernstein, Ashbourn, Sandford  \&
  Allamandola}{Bernstein et~al.}{2004}]{Bernstein2004}
Bernstein M.~P.,  Ashbourn S. F.~M.,  Sandford S.~A.,   Allamandola L.~J.,
  2004, \mn@doi [The Astrophysical Journal] {10.1086/380306}, 601, 365

\bibitem[\protect\citeauthoryear{Bertin, Romanzin, Michaut, Jeseck  \&
  Fillion}{Bertin et~al.}{2011}]{Bertin2011}
Bertin M.,  Romanzin C.,  Michaut X.,  Jeseck P.,   Fillion J.-H.,  2011,
  \mn@doi [The Journal of Physical Chemistry C] {10.1021/jp201487u}, 115, 12920

\bibitem[\protect\citeauthoryear{Blagojevic, Petrie  \& Bohme}{Blagojevic
  et~al.}{2003}]{Blagojevic2003}
Blagojevic V.,  Petrie S.,   Bohme D.~K.,  2003, \mn@doi [Monthly Notices of
  the Royal Astronomical Society] {10.1046/j.1365-8711.2003.06351.x}, 339, L7

\bibitem[\protect\citeauthoryear{Boese, Bläser, Latz  \& Bäumen}{Boese
  et~al.}{1999}]{Boese1999}
Boese R.,  Bläser D.,  Latz R.,   Bäumen A.,  1999, \mn@doi [Acta
  Crystallographica Section C] {https://doi.org/10.1107/S0108270199099862}, 55,
  IUC9900001

\bibitem[\protect\citeauthoryear{Bouilloud, Fray, Bénilan, Cottin, Gazeau  \&
  Jolly}{Bouilloud et~al.}{2015}]{Bouilloud2015}
Bouilloud M.,  Fray N.,  Bénilan Y.,  Cottin H.,  Gazeau M.-C.,   Jolly A.,
  2015, \mn@doi [Monthly Notices of the Royal Astronomical Society]
  {10.1093/mnras/stv1021}, 451, 2145

\bibitem[\protect\citeauthoryear{Burke, Puletti, Woods, Viti, Slater  \&
  Brown}{Burke et~al.}{2015}]{Burke2015}
Burke D.~J.,  Puletti F.,  Woods P.~M.,  Viti S.,  Slater B.,   Brown W.~A.,
  2015, \mn@doi [The Journal of Physical Chemistry A]
  {10.1021/acs.jpca.5b04010}, 119, 6837

\bibitem[\protect\citeauthoryear{Carrascosa, Cruz-Díaz, Muñoz Caro, Dartois
  \& Chen}{Carrascosa et~al.}{2020}]{Carrascosa2020}
Carrascosa H.,  Cruz-Díaz G.~A.,  Muñoz Caro G.~M.,  Dartois E.,   Chen
  Y.-J.,  2020, \mn@doi [Monthly Notices of the Royal Astronomical Society]
  {10.1093/mnras/staa334}, 493, 821

\bibitem[\protect\citeauthoryear{Cazaux, Tielens, Ceccarelli, Castets, Wakelam,
  Caux, Parise  \& Teyssier}{Cazaux et~al.}{2003}]{Cazaux2003}
Cazaux S.,  Tielens A. G. G.~M.,  Ceccarelli C.,  Castets A.,  Wakelam V.,
  Caux E.,  Parise B.,   Teyssier D.,  2003, \mn@doi [The Astrophysical
  Journal] {10.1086/378038}, 593, L51

\bibitem[\protect\citeauthoryear{Clark, Segall, Pickard, Hasnip, Probert,
  Refson  \& Payne}{Clark et~al.}{2005}]{CASTEP}
Clark S.~J.,  Segall M.~D.,  Pickard C.~J.,  Hasnip P.~J.,  Probert M.~J.,
  Refson K.,   Payne M.,  2005, Z. Kristall., 220, 567

\bibitem[\protect\citeauthoryear{Cohen, Todd, Wogan, Black, Keller  \&
  Catling}{Cohen et~al.}{2023}]{Cohen2023}
Cohen Z.~R.,  Todd Z.~R.,  Wogan N.,  Black R.~A.,  Keller S.~L.,   Catling
  D.~C.,  2023, \mn@doi [ACS Earth and Space Chemistry]
  {10.1021/acsearthspacechem.2c00168}, 7, 11

\bibitem[\protect\citeauthoryear{{Crovisier, J.}, {Bockel\'ee-Morvan, D.},
  {Colom, P.}, {Biver, N.}, {Despois, D.}, {Lis, D. C.}  \& {the Team for
  target-of-opportunity radio observations of comets}}{{Crovisier, J.}
  et~al.}{2004}]{Crovisier2004}
{Crovisier, J.} {Bockel\'ee-Morvan, D.} {Colom, P.} {Biver, N.} {Despois, D.}
  {Lis, D. C.}  {the Team for target-of-opportunity radio observations of
  comets} 2004, \mn@doi [A\&A] {10.1051/0004-6361:20035688}, 418, 1141

\bibitem[\protect\citeauthoryear{{Cruz-Diaz}, {Mu\~noz Caro, G. M.}, {Chen,
  Y.-J.}  \& {Yih, T.-S.}}{{Cruz-Diaz} et~al.}{2014}]{Cruz-Diaz2014}
{Cruz-Diaz} G.~A.,  {Mu\~noz Caro, G. M.} {Chen, Y.-J.}  {Yih, T.-S.} 2014,
  \mn@doi [A\&A] {10.1051/0004-6361/201322140}, 562, A119

\bibitem[\protect\citeauthoryear{{D'Hendecourt}, {Allamandola}, {Grim}  \&
  {Greenberg}}{{D'Hendecourt} et~al.}{1986}]{D'Hendecourt1986}
{D'Hendecourt} L.~B.,  {Allamandola} L.~J.,  {Grim} R.~J.~A.,   {Greenberg}
  J.~M.,  1986, \aap, \href
  {https://ui.adsabs.harvard.edu/abs/1986A&A...158..119D} {158, 119}

\bibitem[\protect\citeauthoryear{Ehrenfreund \& Charnley}{Ehrenfreund \&
  Charnley}{2000}]{Ehrenfreund2000}
Ehrenfreund P.,  Charnley S.~B.,  2000, \mn@doi [Annual Review of Astronomy and
  Astrophysics] {10.1146/annurev.astro.38.1.427}, 38, 427

\bibitem[\protect\citeauthoryear{El-Abd, Brogan, Hunter, Willis, Garrod  \&
  McGuire}{El-Abd et~al.}{2019}]{ElAbd2019}
El-Abd S.~J.,  Brogan C.~L.,  Hunter T.~R.,  Willis E.~R.,  Garrod R.~T.,
  McGuire B.~A.,  2019, \mn@doi [The Astrophysical Journal]
  {10.3847/1538-4357/ab3646}, 883, 129

\bibitem[\protect\citeauthoryear{Evans \& Holian}{Evans \&
  Holian}{1985}]{Nose1985}
Evans D.~J.,  Holian B.~L.,  1985, \mn@doi [The Journal of Chemical Physics]
  {10.1063/1.449071}, 83, 4069

\bibitem[\protect\citeauthoryear{{Favre, C.}, {Pagani, L.}, {Goldsmith, P. F.},
  {Bergin, E. A.}, {Carvajal, M.}, {Kleiner, I.}, {Melnick, G.}  \& {Snell,
  R.}}{{Favre, C.} et~al.}{2017}]{Favre2017}
{Favre, C.} {Pagani, L.} {Goldsmith, P. F.} {Bergin, E. A.} {Carvajal, M.}
  {Kleiner, I.} {Melnick, G.}  {Snell, R.} 2017, \mn@doi [A\&A]
  {10.1051/0004-6361/201731327}, 604, L2

\bibitem[\protect\citeauthoryear{Garrison, Haymond, Bennett  \& Cole}{Garrison
  et~al.}{1959}]{Warren1959}
Garrison W.~M.,  Haymond H.~R.,  Bennett W.,   Cole S.,  1959, Radiation
  Research, 10, 273

\bibitem[\protect\citeauthoryear{Garrod}{Garrod}{2013}]{Garrod2013}
Garrod R.~T.,  2013, \mn@doi [The Astrophysical Journal]
  {10.1088/0004-637X/765/1/60}, 765, 60

\bibitem[\protect\citeauthoryear{{Gerakines}, {Schutte}, {Greenberg}  \& {van
  Dishoeck}}{{Gerakines} et~al.}{1995}]{Gerakines1995}
{Gerakines} P.~A.,  {Schutte} W.~A.,  {Greenberg} J.~M.,   {van Dishoeck}
  E.~F.,  1995, \aap, \href
  {http://adsabs.harvard.edu/abs/1995A%26A...296..810G} {296, 810}

\bibitem[\protect\citeauthoryear{Godfrey, Mirabella  \& Brown}{Godfrey
  et~al.}{2000}]{Godfrey2000}
Godfrey P.~D.,  Mirabella M.~J.,   Brown R.~D.,  2000, \mn@doi [The Journal of
  Physical Chemistry A] {10.1021/jp992499t}, 104, 258

\bibitem[\protect\citeauthoryear{Gonz\'alez~D\'iaz, Carrascosa, Mu\~noz Caro,
  Satorre  \& Chen}{Gonz\'alez~D\'iaz et~al.}{2022}]{GonzalezDiaz2022}
Gonz\'alez~D\'iaz C.,  Carrascosa H.,  Mu\~noz Caro G.~M.,  Satorre M.~A.,
  Chen Y.-J.,  2022, \mn@doi [Monthly Notices of the Royal Astronomical
  Society] {10.1093/mnras/stac3122}, 517, 5744

\bibitem[\protect\citeauthoryear{Hagen, Tielens  \& Greenberg}{Hagen
  et~al.}{1981}]{HAGEN1981}
Hagen W.,  Tielens A.,   Greenberg J.,  1981, \mn@doi [Chemical Physics]
  {https://doi.org/10.1016/0301-0104(81)80158-9}, 56, 367

\bibitem[\protect\citeauthoryear{Hohenberg \& Kohn}{Hohenberg \&
  Kohn}{1964}]{DFT-HK}
Hohenberg P.,  Kohn W.,  1964, Phys. Rev., 136, B864

\bibitem[\protect\citeauthoryear{{Huntress} \& {Mitchell}}{{Huntress} \&
  {Mitchell}}{1979}]{Huntress1979}
{Huntress} W.~T. J.,  {Mitchell} G.~F.,  1979, \mn@doi [\apj] {10.1086/157207},
  \href {https://ui.adsabs.harvard.edu/abs/1979ApJ...231..456H} {231, 456}

\bibitem[\protect\citeauthoryear{Jackson, Stibrich, McLain, Fondren, Adams  \&
  Babcock}{Jackson et~al.}{2005}]{jackson2005}
Jackson D.~M.,  Stibrich N.~J.,  McLain J.~L.,  Fondren L.~D.,  Adams N.~G.,
  Babcock L.~M.,  2005, \mn@doi [International Journal of Mass Spectrometry]
  {https://doi.org/10.1016/j.ijms.2005.09.001}, 247, 55

\bibitem[\protect\citeauthoryear{Kleimeier, Eckhardt  \& Kaiser}{Kleimeier
  et~al.}{2020}]{Kleimeier2020}
Kleimeier N.~F.,  Eckhardt A.~K.,   Kaiser R.~I.,  2020, \mn@doi [The
  Astrophysical Journal] {10.3847/1538-4357/abafa4}, 901, 84

\bibitem[\protect\citeauthoryear{Kohn \& Sham}{Kohn \& Sham}{1965}]{DFT-KS}
Kohn W.,  Sham L.~J.,  1965, Phys. Rev., 140, A1133

\bibitem[\protect\citeauthoryear{Krause, Katon, Rogers  \& Phillips}{Krause
  et~al.}{1977}]{Krause_Katon1977}
Krause P.~F.,  Katon J.~E.,  Rogers J.~M.,   Phillips D.~B.,  1977, \mn@doi
  [Applied Spectroscopy] {10.1366/000370277774463959}, 31, 110

\bibitem[\protect\citeauthoryear{Lafosse, Bertin, Domaracka, Pliszka,
  Illenberger  \& Azria}{Lafosse et~al.}{2006}]{Lafosse2006}
Lafosse A.,  Bertin M.,  Domaracka A.,  Pliszka D.,  Illenberger E.,   Azria
  R.,  2006, \mn@doi [Phys. Chem. Chem. Phys.] {10.1039/B613479C}, 8, 5564

\bibitem[\protect\citeauthoryear{{Lattelais} et~al.,}{{Lattelais}
  et~al.}{2011}]{Lattelais2011}
{Lattelais} et~al., 2011, \mn@doi [A\&A] {10.1051/0004-6361/201016184}, 532,
  A12

\bibitem[\protect\citeauthoryear{Leach, Schwell, Un, Jochims  \&
  Baumgärtel}{Leach et~al.}{2006}]{Leach2006}
Leach S.,  Schwell M.,  Un S.,  Jochims H.-W.,   Baumgärtel H.,  2006, \mn@doi
  [Chemical Physics] {https://doi.org/10.1016/j.chemphys.2005.08.044}, 321, 159

\bibitem[\protect\citeauthoryear{Lopes, Domanskaya, Fausto, Räsänen  \&
  Khriachtchev}{Lopes et~al.}{2010}]{Lopes2010}
Lopes S.,  Domanskaya A.~V.,  Fausto R.,  Räsänen M.,   Khriachtchev L.,
  2010, \mn@doi [The Journal of Chemical Physics] {10.1063/1.3484943}, 133

\bibitem[\protect\citeauthoryear{{Manigand} et~al.,}{{Manigand}
  et~al.}{2020}]{manigand2020}
{Manigand} S.,  et~al., 2020, \mn@doi [\aap] {10.1051/0004-6361/201936299},
  \href {https://ui.adsabs.harvard.edu/abs/2020A&A...635A..48M} {635, A48}

\bibitem[\protect\citeauthoryear{{Mart\'{\i}n-Dom\'enech, R.,} \& {Goesmann,
  F.}}{{Mart\'{\i}n-Dom\'enech, R.,} \& {Goesmann,
  F.}}{2014}]{Martin-Domenech2014}
{Mart\'{\i}n-Dom\'enech, R.,} {Mu\~noz Caro, G. M.} B.,  {Goesmann, F.} 2014,
  \mn@doi [A\&A] {10.1051/0004-6361/201322824}, 564, A8

\bibitem[\protect\citeauthoryear{{Mart\'{\i}n-Dom\'enech, R.,},
  {Manzano-Santamar\'{\i}a, J.}, {Mu\~noz Caro, G. M.}, {Cruz-D\'{\i}az, G.
  A.}, {Chen, Y.-J.}, {Herrero, V. J.}  \& {Tanarro,
  I.}}{{Mart\'{\i}n-Dom\'enech, R.,} et~al.}{2015}]{Martin-Domenech2015}
{Mart\'{\i}n-Dom\'enech, R.,} {Manzano-Santamar\'{\i}a, J.} {Mu\~noz Caro, G.
  M.} {Cruz-D\'{\i}az, G. A.} {Chen, Y.-J.} {Herrero, V. J.}  {Tanarro, I.}
  2015, \mn@doi [A\&A] {10.1051/0004-6361/201526003}, 584, A14

\bibitem[\protect\citeauthoryear{Maçôas, Khriachtchev, Fausto  \&
  Räsänen}{Maçôas et~al.}{2004}]{Macoas2004}
Maçôas E. M.~S.,  Khriachtchev L.,  Fausto R.,   Räsänen M.,  2004, \mn@doi
  [The Journal of Physical Chemistry A] {10.1021/jp037840v}, 108, 3380

\bibitem[\protect\citeauthoryear{McGuire}{McGuire}{2022}]{McGuire2022}
McGuire B.~A.,  2022, \mn@doi [The Astrophysical Journal Supplement Series]
  {10.3847/1538-4365/ac2a48}, 259, 30

\bibitem[\protect\citeauthoryear{{Mehringer}, {Snyder}, {Miao}  \&
  {Lovas}}{{Mehringer} et~al.}{1997}]{Mehringer1997}
{Mehringer} D.~M.,  {Snyder} L.~E.,  {Miao} Y.,   {Lovas} F.~J.,  1997, \mn@doi
  [\apjl] {10.1086/310612}, \href
  {https://ui.adsabs.harvard.edu/abs/1997ApJ...480L..71M} {480, L71}

\bibitem[\protect\citeauthoryear{{Modica} \& {Palumbo}}{{Modica} \&
  {Palumbo}}{2010}]{Modica2010}
{Modica} P.,  {Palumbo} M.~E.,  2010, \mn@doi [\aap]
  {10.1051/0004-6361/201014101}, \href
  {https://ui.adsabs.harvard.edu/abs/2010A&A...519A..22M} {519, A22}

\bibitem[\protect\citeauthoryear{{Mu\~noz Caro}, {Jim\'enez-Escobar},
  {Mart\'{\i}n-Gago}, {Rogero}, {Atienza}, {Puertas}, {Sobrado}  \&
  {Torres-Redondo}}{{Mu\~noz Caro} et~al.}{2010}]{MuñozCaro2010}
{Mu\~noz Caro} G.~M.,  {Jim\'enez-Escobar} A.,  {Mart\'{\i}n-Gago} J.~A.,
  {Rogero} C.,  {Atienza} C.,  {Puertas} S.,  {Sobrado} J.~M.,
  {Torres-Redondo} J.,  2010, \mn@doi [A\&A] {10.1051/0004-6361/200912462},
  522, A108

\bibitem[\protect\citeauthoryear{Muthuselvi, Arunkumar  \&
  Rajaperumal}{Muthuselvi et~al.}{2016}]{Muthuselvi2016}
Muthuselvi C.,  Arunkumar A.,   Rajaperumal G.,  2016, Pelagia Research Library
  Der Chemica Sinica,, 7

\bibitem[\protect\citeauthoryear{{Noriega-Crespo} et~al.,}{{Noriega-Crespo}
  et~al.}{2004}]{Noriega2004}
{Noriega-Crespo} A.,  et~al., 2004, \mn@doi [\apjs] {10.1086/422819}, \href
  {https://ui.adsabs.harvard.edu/abs/2004ApJS..154..352N} {154, 352}

\bibitem[\protect\citeauthoryear{Perdew, Burke  \& Ernzerhof}{Perdew
  et~al.}{1996}]{PBE1996}
Perdew J.~P.,  Burke K.,   Ernzerhof M.,  1996, \mn@doi [Phys. Rev. Lett.]
  {10.1103/PhysRevLett.77.3865}, 77, 3865

\bibitem[\protect\citeauthoryear{{Pilling}, {Santos, A. C. F.}  \&
  {Boechat-Roberty, H. M.}}{{Pilling} et~al.}{2006}]{Pilling2006}
{Pilling} {Santos, A. C. F.}  {Boechat-Roberty, H. M.} 2006, \mn@doi [A\&A]
  {10.1051/0004-6361:20053927}, 449, 1289

\bibitem[\protect\citeauthoryear{Rachid, Faquine  \& Pilling}{Rachid
  et~al.}{2017}]{RACHID2017}
Rachid M.~G.,  Faquine K.,   Pilling S.,  2017, \mn@doi [Planetary and Space
  Science] {https://doi.org/10.1016/j.pss.2017.05.003}, 149, 83

\bibitem[\protect\citeauthoryear{Refson, Tulip  \& Clark}{Refson
  et~al.}{2006}]{CASTEP_DFPT}
Refson K.,  Tulip P.~R.,   Clark S.~J.,  2006, Phys. Rev. B, 73, 155114

\bibitem[\protect\citeauthoryear{Remijan, Snyder, Liu, Mehringer  \&
  Kuan}{Remijan et~al.}{2002}]{Remijan2002}
Remijan A.,  Snyder L.~E.,  Liu S.-Y.,  Mehringer D.,   Kuan Y.-J.,  2002,
  \mn@doi [The Astrophysical Journal] {10.1086/341627}, 576, 264

\bibitem[\protect\citeauthoryear{Remijan, Snyder, Friedel, Liu  \&
  Shah}{Remijan et~al.}{2003}]{Remijan2003}
Remijan A.,  Snyder L.~E.,  Friedel D.~N.,  Liu S.-Y.,   Shah R.~Y.,  2003,
  \mn@doi [The Astrophysical Journal] {10.1086/374890}, 590, 314

\bibitem[\protect\citeauthoryear{Remijan, Wyrowski, Friedel, Meier  \&
  Snyder}{Remijan et~al.}{2005}]{Remijan2005}
Remijan A.~J.,  Wyrowski F.,  Friedel D.~N.,  Meier D.~S.,   Snyder L.~E.,
  2005, \mn@doi [The Astrophysical Journal] {10.1086/429750}, 626, 233

\bibitem[\protect\citeauthoryear{Sanz-Novo et~al.,}{Sanz-Novo
  et~al.}{2023}]{SanzNovo2023}
Sanz-Novo M.,  et~al., 2023, \mn@doi [The Astrophysical Journal]
  {10.3847/1538-4357/ace523}, 954, 3

\bibitem[\protect\citeauthoryear{Sato et~al.,}{Sato et~al.}{2018}]{Sato2018}
Sato A.,  et~al., 2018, \mn@doi [Molecular Astrophysics]
  {https://doi.org/10.1016/j.molap.2018.01.002}, 10, 11

\bibitem[\protect\citeauthoryear{Schuhmann et~al.,}{Schuhmann
  et~al.}{2019}]{Schuhmann2019}
Schuhmann M.,  et~al., 2019, \mn@doi [ACS Earth and Space Chemistry]
  {10.1021/acsearthspacechem.9b00094}, 3, 1854

\bibitem[\protect\citeauthoryear{{Schutte} et~al.,}{{Schutte}
  et~al.}{1999}]{Schutte1999}
{Schutte} W.~A.,  et~al., 1999, \aap, \href
  {https://ui.adsabs.harvard.edu/abs/1999A&A...343..966S} {343, 966}

\bibitem[\protect\citeauthoryear{Shiao, Looney, Remijan, Snyder  \&
  Friedel}{Shiao et~al.}{2010}]{Shiao2010}
Shiao Y.-S.~J.,  Looney L.~W.,  Remijan A.~J.,  Snyder L.~E.,   Friedel D.~N.,
  2010, \mn@doi [The Astrophysical Journal] {10.1088/0004-637X/716/1/286}, 716,
  286

\bibitem[\protect\citeauthoryear{Yan \& Chu}{Yan \& Chu}{2008}]{Huiyan2008}
Yan H.,  Chu L.~T.,  2008, \mn@doi [Langmuir] {10.1021/la8008706}, 24, 9410

\bibitem[\protect\citeauthoryear{Öberg, van Dishoeck, Linnartz  \&
  Andersson}{Öberg et~al.}{2010}]{Oberg2010}
Öberg K.~I.,  van Dishoeck E.~F.,  Linnartz H.,   Andersson S.,  2010, \mn@doi
  [The Astrophysical Journal] {10.1088/0004-637X/718/2/832}, 718, 832

\makeatother
\end{thebibliography}

% Don't change these lines
\bsp	% typesetting comment
\label{lastpage}

%\begin{thebibliography}{99}
%\end{thebibliography}

%%%%%%%%%%%%%%%%%%%%%%%%%%%%%%%%%%%%%%%%%%%%%%%%%%

%%%%%%%%%%%%%%%%% APPENDICES %%%%%%%%%%%%%%%%%%%%%

%\appendix



%%%%%%%%%%%%%%%%%%%%%%%%%%%%%%%%%%%%%%%%%%%%%%%%%%
\end{document}